\documentclass[%
 aip,
 amsmath,amssymb
 reprint,%
]{revtex4-1}
\usepackage{graphicx}
\usepackage{dcolumn}
\usepackage{bm}
\usepackage{amssymb}
\usepackage[dvipsnames]{xcolor}
\usepackage{soul}
\usepackage[utf8]{inputenc}
\usepackage[T1]{fontenc}
\usepackage{mathptmx}
\usepackage{etoolbox}


\begin{document}


\title{Effect of thermal-bioconvection in a rotating phototactic medium}
\author{Sandeep Kumar}
\email{sandeepkumar1.00123@gmail.com}
\affiliation{
Department of Mathematics, PDPM  Indian Institute of Information Technology Design and Manufacturing, Jabalpur 482005, India
}%
\author{Shaowei Wang}
\email{shaoweiwang@sdu.edu.cn}
\affiliation{
Department of Engineering Mechanics, School of Civil Engineering, Shandong University, Jinan 250061, PR China
}


\date{\today}

\begin{abstract}

In this article, thermal bioconvection in a rotating phototactic medium is analyzed with a stress-free top and rigid upper boundary. The suspension rotates with a uniform angular velocity around a vertical axis.
Utilizing MATLAB's bvp$4$c solver, neutral and growth rate curves are analyzed, emphasizing the impacts of parameters such as the Taylor number, thermal Rayleigh number, Lewis number, and critical total intensity. It is observed that generally critical bioconvection Rayleigh number increases with increasing Taylor number and Lewis number while decreasing with higher thermal Rayleigh number and critical total intensity. The critical thermal Rayleigh number appears to remain unaffected by variations in the critical total intensity and Lewis number. However, it is notably influenced by changes in the Taylor number.

\end{abstract}

\maketitle

\begin{figure*}
    \centering
    \includegraphics[width=16cm, height=10cm ]{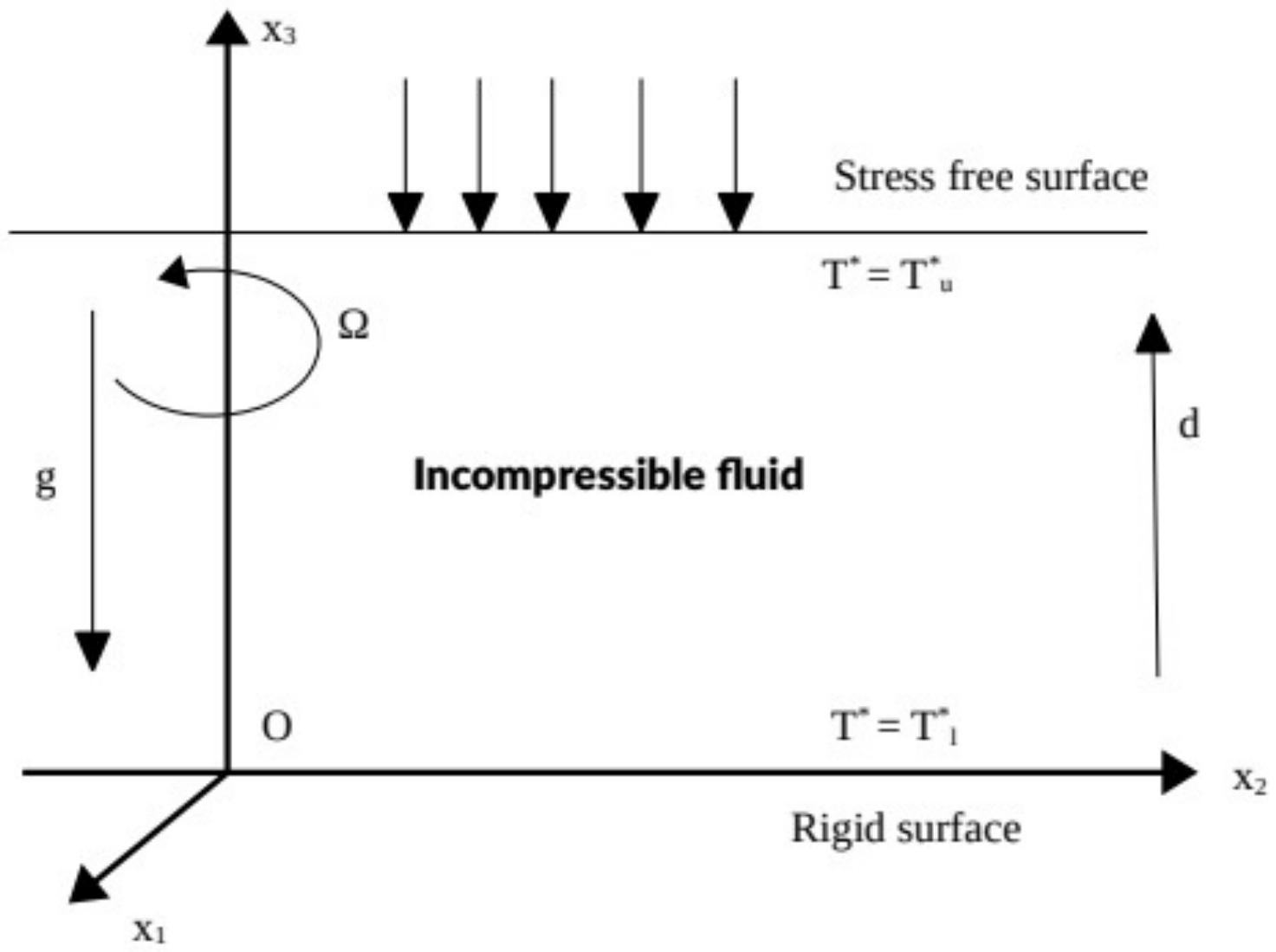}
    \caption{The configuration of the problem.}
   \label{fig:profile-flow.eps}
 \end{figure*}

\section{Introduction}

The phenomenon of bioconvection involves the emergence of patterns generated by microorganisms in suspensions, with their movements not predetermined \cite{ref1}. The early experiments and the term "bioconvection" were introduced by the pioneering work of \citet{ref2} and \citet{ref3} respectively. These microorganisms, slightly denser than the surrounding fluid, tend to aggregate due to tactic behavior. Notably, their upward movement contributes to pattern formation, while the absence of such movement renders these patterns invisible. It is essential to note that the specific types of patterns formed can be influenced by parameters such as depth, microorganism concentration, and motion. Several microorganisms, including flagellated green algae such as \textit{Volvox}, \textit{Euglena}, \textit{Dunaliella}, and \textit{Chlamydomonas}, demonstrate the formation of these patterns \cite{ref6,ref5,ref4,ref2}.

The collective responses of microorganisms to external stimuli are known as taxes, encompassing various types such as chemotaxis, gravitaxis, gyrotaxis, and phototaxis. Chemotaxis involves responses to changes in chemical gradients, while gravitaxis pertains to reactions influenced by alterations in the force of gravity. The balance between the torque induced by gravity and shear results in gyrotaxis. Phototaxis, as discussed in this context, encompasses movements toward or away from light. Experiments have shed light on the roles of light intensity and gradient in determining bioconvection patterns, with variations in pattern contour, size, and scale linked to the amount of light \cite{ref5,ref2,ref7}. The interplay between the total light intensity $\mathcal{G}$ and a critical value $\mathcal{G}_c$, alongside the light absorption by microorganisms, contributes significantly to pattern variations \cite{ref8}.

Within a finite-depth suspension that is illuminated from above, the equilibrium state is established through a delicate balance between phototaxis induced by diffusion and light absorption. This equilibrium gives rise to a horizontally oriented sublayer densely populated by microorganisms. Notably, the region above this sublayer maintains its gravitational stability, unlike the region below it, which undergoes gravitational instability. The pivotal parameter, denoted as $\mathcal{G}_c$, influences the sublayer's location. When the total intensity $\mathcal{G}$ across the entire suspension is less than $\mathcal{G}_c$, the sublayer forms at the upper boundary. Conversely, if $\mathcal{G}$ exceeds this threshold, the sublayer emerges at the lower boundary. When the critical total intensity $\mathcal{G}_c$ matches the overall intensity $\mathcal{G}$, the sublayer positions itself midway between the upper and lower bounds. In the event that an unstable fluid layer arises, its movement extends into the stable layer through a process known as penetrative convection \cite{ref10}.

The study of fluids in rotation introduces a captivating dimension to fluid dynamics research. Various researchers have delved into the complexities of fluid dynamics in the presence of rotation, investigating different aspects such as convection, instability, and the impact of external forces like magnetic fields. In the context of rotating viscous fluids, the B\'{e}nard aspect has been extensively explored, particularly in the presence of rotational effects \citep{ref11, ref12}. Researchers have extended their investigations to include bioconvection in a rotating medium, exploring phenomena such as thermo-convective instability and the influence of magnetic fields aligned along the vertical axis \citep{ref12-1, ref12-2}. While rotation can stabilize the system, it does so at a slower rate compared to non-rotating viscous fluids. Various studies have employed linear and weakly non-linear analysis of the Boussinesq approximation's governing hydrodynamic equations to understand the impact of rotation on fluid instabilities \citep{ref12-3, ref12-4}. These studies have shown that the behaviors of instabilities are closely tied to the boundary conditions, particularly at the bottom surface. 
Furthermore, researchers have focused on specific fluid types, such as nanofluids, examining their thermal instability and the effects of rotation on convective patterns \citep{ref12-5, ref12-6}. The findings revealed that factors such as magnetic field strength, rotation speed, and particle concentration play a critical role in determining the stability or instability of the system, especially in the vicinity of the lower boundary. Studies have also analyzed the flow characteristics of various types of nanofluids, including Reiner-Rivlin and Casson nanofluids, in the presence of rotation, incorporating factors like thermal radiation, gyrotactic microorganisms, and multiple slips \citep{ref12-7, waqas}. These investigations collectively contribute to a better understanding of the intricate dynamics of rotating fluid systems and their diverse applications.

Early studies primarily focused on suspensions in fluids at constant temperatures, yet many microbes, particularly thermophilic ones found in hot springs, thrive in environments characterized by wide temperature variations. Several authors have delved into the realm of biothermal convection. For instance, \citet{kuznetsov2005onset} explored the instability of gyrotactic microorganisms subjected to heating from below in suspensions, revealing a connection between bioconvection and natural convection through linear stability analysis. Furthermore, \citet{kuznetsov2011non} delved into biothermal convection in suspensions comprising both gyrotactic microorganisms and nanoparticles, investigating both stationary and oscillatory convection and examining thermal and bioconvection Rayleigh numbers. In a related vein, \citet{zhao2018linear} employed a random swimming model to investigate the linear stability of gyrotactic microorganisms subjected to heating from below, uncovering non-oscillatory instability under specific conditions and rigid boundary constraints. Using the Darcy–Brinkman model, \citet{zhao2019darcy} probed bio-thermal convection within a highly porous medium, where heat was applied from below. Additionally, \citet{balla2020bioconvection} explored bioconvection in a square enclosure with porous walls, with a focus on oxytactic microorganisms influenced by thermal radiation. In the context of a vertical wavy porous cavity, \citet{hussain2022thermal} investigated the impact of thermal radiation on the bioconvection flow of nano-enhanced phase change materials and oxytactic microorganisms.

Initially, phototactic bioconvection was introduced by \citet{ref9}, who examined a model for non-scattering absorbing suspensions. This model considers the stability of suspensions of phototactic microorganisms using a phototaxis and shading model. The suspension is uniformly illuminated from above, with microorganisms swimming in a slightly less dense fluid. Subsequently, \citet{ref13} introduced a two-dimensional phototactic model and investigated its stability, using a finite-difference approach for numerical solution. However, these early models did not account for scattering effects. \citet{ref14} extended the phototactic bioconvection model to include isotropic scattering, revealing that under specific parameter values, microorganisms congregate in two distinct horizontal layers at different depths due to scattering. An extension of this work by \citet{kumar2023} demonstrated that the system exhibited greater stability when the top surface was rigid compared to a stress-free surface. The impact of forward anisotropic scattering on phototactic bioconvection suspensions was examined by \citet{ref15}, with a primary focus on forward scattering effects. \citet{ref16} and \citet{panda2020effects} further explored how scattering suspensions were influenced by both diffuse and collimated light, demonstrating a diverse range of behaviors under varying conditions. \citet{ref17} investigated a phototactic bioconvection model under oblique light and in a non-scattering suspension, observing transitions between stable and overstable modes with different angles of incidence and specific parameter variations. The effect of oblique irradiation on an isotropic scattering suspension was studied by \citet{ref18} and \citet{panda2023phototactic}, revealing distinct concentration profile patterns for varied scattering albedo values and angles of incidence. Notably, none of these phototactic bioconvection models considered the influence of rotation. Recently, \citet{kumar2023effect} investigated the effect of rotation on non-scattering media and analyzed linear instability, illustrating the stabilizing effect of rotation on the suspension. Despite these intriguing findings, there has been no theoretical exploration considering both phototactic and temperature gradient processes.

In this present study, we employ the phototaxis model introduced by \citet{kumar2023effect}, utilizing the Navier-Stokes equation in conjunction with a cell conservation equation to accurately model an incompressible fluid. Our phototactic suspension, under the proposed hypothesis, is heated from below and illuminated from above via vertical collimated irradiation. Given the dependence of a significant number of motile algae on photosynthesis for their nutrition, an investigation of a phototaxis model within a thermal medium is essential to present this model realistically.

The structure of this paper unfolds as follows: Section \ref{sec2} presents the mathematical model for phototactic bioconvection, along with corresponding boundary conditions. The steady-state solution is explored in Section \ref{sec3}, followed by a linear stability analysis in Section \ref{sec4}. Subsequently, Section \ref{sec5} explains the numerical results, while the conclusion is discussed in Section \ref{sec6}.

\section{MATHEMATICAL FORMULATION}
\label{sec2}
Consider a circumstance in which a fluid and plates are in angular motion, rotating around the $x_3$-axis with a constant angular velocity $\boldsymbol{\Omega}=\Omega \boldsymbol{k}$ in a finite-depth Cartesian coordinate system $O x_1 x_2 x_3$. The lower boundary at $x_3 = 0$ is assumed to be rigid, whereas the top boundary at $x_3 = d$ is supposed to be stress-free. Vertical, collimated irradiation from a light source illuminates the upper border of the suspension, and the effects of oblique or diffused irradiation are ignored. Also, isotropic/anisotropic scattering is ignored, considering only light absorption in the medium.



The continuity equation is given by
\begin{equation}
\label{eqn:equation8}
    div(\boldsymbol{v})=0.
\end{equation}

The equation for momentum in the rotating medium, related to the Boussinesq, can be written as  \cite{ref12}

\begin{align}
    \label{eqn:equation9}
    \varrho\left(\frac{\partial}{\partial t}+\boldsymbol{v}\cdot\nabla\right)\boldsymbol{v}+2\boldsymbol{\Omega}\times\boldsymbol{v}&=\mu\nabla^2\boldsymbol{v}-\boldsymbol{\nabla} \mathcal{P}-n\vartheta \Delta \varrho g \boldsymbol{k} 
    -\varrho (1-\beta(T^*-T_u^*))g \boldsymbol{k}.
\end{align}

In this context, time is denoted by $t$, fluid velocity by $\boldsymbol{v}$, dynamic viscosity by $\mu$, fluid density by $\rho$, the density of each algal cell by $\rho+\Delta \rho$ $(\Delta \rho/\rho \ll 1)$, volume of each algal cell by $\vartheta$, gravitational acceleration by $g$, cell concentration by $n$, excess pressure above hydrostatic by $\mathcal{P}$, volumetric thermal expansion coefficient by $\beta$. The temperature of the fluid is indicated by $T^*$, while the temperature of the upper wall is represented by $T_u^*$ and the temperature of the lower wall is represented by $T_l^*$ accordingly.  The rotational effect is expressed only by the Coriolis force, which is the last component in equation (\ref{eqn:equation9}) on the left-hand side.

The thermal energy equation is given by
\begin{equation}
    \varrho c \left[\frac{\partial T^*}{\partial t}+ \boldsymbol{v}\cdot\boldsymbol{\nabla} T^*\right]=\boldsymbol{\nabla}(k \boldsymbol{\nabla} T^*).
\end{equation}

Here, the thermal conductivity and the volumetric heat capacity of water are denoted by $k$ and $\varrho c$, respectively.

The equation for the conservation of cells can be written as follows \cite{ref1,ref9}:
\begin{eqnarray}
    \label{eqn:equation10}
    \frac{\partial n}{\partial t}=-\boldsymbol{\nabla}\cdot\boldsymbol{A}, \nonumber\\ 
    \boldsymbol{A}=n \boldsymbol{U}_c+n\boldsymbol{v}-\boldsymbol{D}\cdot\boldsymbol{\nabla} n,
\end{eqnarray}
where $\boldsymbol{A}$ represents the flux of cells and is defined as the sum of three components: the mean cells' swimming velocity, the flux caused by the advection of the cells, and the random component of the cell's locomotion. $\boldsymbol{D}=D\boldsymbol{I}$ depicts the isotropic and constant diffusivity tensor where $\boldsymbol{I}$ representing the identity tensor and $D$ is the diffusion coefficient.
Maintaining mass and cell conservation in an incompressible fluid suspension requires horizontal boundaries. No fluid or cell movement across these boundaries is required to preserve these concepts. The vertical velocity and cell movement at these borders must be zero. The fluid velocity at these boundaries depends on the type of surface. One makes a distinction between a stress-free boundary and a rigid one. Tangential stress must be zero for a stress-free boundary. This stress-free surface permits the fluid to flow parallel to the boundary. Rigid boundaries have a distinct impact, it inhibits fluid flow. This requires all velocity components to be zero. The fluid can’t move along a rigid boundary, therefore its velocity in all directions must disappear.
The boundary conditions for the stress-free top surface
\begin{align}
\label{eqn:equation12}
    \boldsymbol{v}\cdot\boldsymbol{k}=0, \quad \frac{\partial^2}{\partial x_3^2}(\boldsymbol{v}\cdot\boldsymbol{k})=0, \quad T^*&=T_u^*, \quad \boldsymbol{A}\cdot\boldsymbol{k}=0,  \quad \nonumber\\
    &\text{at} \quad x_3=H,
\end{align}
while for the rigid bottom surface 
\begin{equation}
\label{eqn:equation11}
    \boldsymbol{v}=0, \quad \boldsymbol{v}\times\boldsymbol{k}=0, \quad  T^*=T_l^*, \quad\boldsymbol{A}\cdot\boldsymbol{k}=0,\quad \text{at} \quad x_3=0.
\end{equation}

The radiation transfer equation \cite{ref-modest,ref-chand}
\begin{equation}
\label{eqn:equation3}
    \boldsymbol{r}\cdot \boldsymbol{\nabla}I(\boldsymbol{x},\boldsymbol{r})+\psi I(\boldsymbol{x},\boldsymbol{r})=0,
\end{equation}
is used to calculate the intensity $I(\boldsymbol{x},\boldsymbol{r})$ of a light beam passing through an absorbing medium that does not scatter light. In this context, the absorption coefficient $\psi$ is defined with the concentration $n$ by a linear proportionality, which can be written as $\psi=\iota n$, where $\iota$ is the proportionality constant.
The boundary conditions are such that light reflection on the top and bottom surfaces is not considered. Therefore, the top surface condition can be written as
\begin{equation}
\label{eqn:equation4}
    I(x_1,x_2,H,\alpha_1,\alpha_2)=I^0\delta(\boldsymbol{r}-\boldsymbol{r}^0) , \quad \pi/2\le\alpha_1\le\pi,
\end{equation}
and the bottom surface condition as 
\begin{equation}
    I(x_1,x_2,0,\alpha_1,\alpha_2)=0,\quad 0\le\alpha_1\le\pi/2.
\end{equation}

The incident direction is denoted by $\boldsymbol{r}^0=-\boldsymbol{k}$, the incident irradiation is given by $I^0$, and the Dirac delta function is subject to the condition \cite{ref-modest}
\begin{equation*}
    \int_0^{4\pi} f(\boldsymbol{r})\delta(\boldsymbol{r}-\boldsymbol{r}^0) d\omega=f(\boldsymbol{r}^0).
\end{equation*}

At a point $\boldsymbol{x}=(x_1,x_2,x_3)$ the radiative heat flux $\boldsymbol{q}(\boldsymbol{x})$ and the total intensity $\mathcal{G}(\boldsymbol{x})$ are given as \cite{ref-modest}
\begin{equation}
\label{eqn:equation2}
    \boldsymbol{q}(\boldsymbol{x})=\int_0^{4\pi}I(\boldsymbol{x},\boldsymbol{r})\boldsymbol{r}d\omega,
\end{equation}
\begin{equation}
\label{eqn:equation1}
    \mathcal{G}(\boldsymbol{x})=\int_0^{4\pi}I(\boldsymbol{x},\boldsymbol{r})d\omega.
\end{equation}
Here the solid angle is indicated by $\omega$.

The average cell swimming velocity is 
\begin{equation}
\label{eqn:equation5}
    \boldsymbol{U}_c=U_c\bar{\boldsymbol{P}},
\end{equation}
where $U_c$ is the average cell swimming speed and $\bar{\boldsymbol{P}}$ is the average cell swimming direction which is calculated as \cite{ref9}
\begin{equation}
\label{eqn:equation6}
    \bar{\boldsymbol{P}}=T(\mathcal{G})\boldsymbol{k}.
\end{equation}

The formulation of taxis function $T(\mathcal{G})$ is defined as follows:
\begin{equation}
\label{eqn:equation7}
 T(\mathcal{G})  \left\{ \begin{array}{lll}
 \ge 0 & \mbox{for $\mathcal{G}_c \ge \mathcal{G}$};\\
         < 0 & \mbox{for $\mathcal{G}_c < \mathcal{G}$},
        \end{array} \right. 
\end{equation}
The precise expression of the phototaxis function $T(\mathcal{G})$, varies among microorganisms due to their distinctive characteristics \cite{ref9}. To provide a further level of understanding of this concept, a particular implementation of the phototaxis function might be described as follows: \cite{ref13,ref17}
\begin{equation*}
    T(\mathcal{G})=0.8\sin{\left(\frac{3}{2}\pi\varphi(\mathcal{G})\right)}-0.1\sin{\left(\frac{1}{2}\pi\varphi(\mathcal{G})\right)},
\end{equation*}
where  $\varphi=\mathcal{G}e^{\chi(\mathcal{G}-1)}$. The parameter  $\chi$ depends on the critical total intensity $\mathcal{G}_c$.

To construct the non-dimensional bioconvection equations, one must first scale all of the necessary factors, including lengths, pressure, fluid velocity, time, and cell concentration. This scaling is accomplished by the use of standard parameters including $H$, $\mu \alpha_f/H^{2}$, $\alpha_f/H$, $H^{2}/\alpha_f$, and $\bar{n}$. For the temperature, the non-dimensionalization is accomplished by defining $\mathcal{T}=\frac{T^*-T_u^*}{\Delta T}$, where $\alpha_f$ stands for the thermal diffusivity of water and $\Delta T= T_l^*-T_h^*$, which is the temperature difference that occurs between the upper and lower borders.

After performing the required changes to the governing equations (\ref{eqn:equation8})-(\ref{eqn:equation10}) by replacing corresponding dimensionless variables:
\begin{equation}
\label{eqn:equation13}
    div(\boldsymbol{v})=0, 
\end{equation}
\begin{equation}
\label{eqn:equation14}
    \frac{1}{P_r}\left(\frac{\partial}{\partial t}+\boldsymbol{v}\cdot\boldsymbol{\nabla}\right)\boldsymbol{v}+Ta^{1/2}(\boldsymbol{k}\times\boldsymbol{v})=\nabla^2\boldsymbol{v}-\boldsymbol{\nabla} \mathcal{P}-n R_a \boldsymbol{k}-R_m \boldsymbol{k}+R_T \mathcal{T} \boldsymbol{k},
\end{equation}
\begin{equation}
    \frac{\partial \mathcal{T} }{\partial t}+ \boldsymbol{v}\cdot \boldsymbol{\nabla} \mathcal{T}=\boldsymbol{\nabla}^2 \mathcal{T},
\end{equation}
\begin{equation}
    \label{eqn:equation15}
    \frac{\partial n}{\partial t}=-\boldsymbol{\nabla}\cdot\left(\frac{1}{Le}n U_s\bar{\boldsymbol{P}}+n\boldsymbol{v}-\frac{1}{Le}\boldsymbol{\nabla} n\right).
\end{equation}

Here, the basic-density Rayleigh number is $R_m=\frac{\varrho g H^3}{\mu \alpha_f}$, the bioconvection Rayleigh number is $R_a=\frac{\bar{n}\vartheta \Delta \rho g H^3}{\mu \alpha_f}$, the Taylor number is $Ta=\frac{4\Omega^2 H^4}{\nu^2}$, the thermal Rayleigh number is $R_T=\frac{\beta \Delta T \varrho g H^3}{\mu \alpha_f}$, the Lewis number is $Le=\frac{\alpha_f}{D}$, the Prandtl number is $P_r=\frac{\mu}{\varrho \alpha_f}$, dimensionless swimming speed is $U_s=\frac{U_c H}{D}$, and the kinematic viscosity is $\nu=\mu/\rho$.

Non-dimensional boundary conditions for the stress-free top surface become
\begin{align}
\label{eqn:equation17}
    \boldsymbol{v}\cdot\boldsymbol{k}&=0, \quad \frac{\partial^2}{\partial x_3^2}(\boldsymbol{v}.\boldsymbol{k})=0, \quad \mathcal{T}=0, \quad \nonumber \\
    &\left(\frac{1}{Le}n U_s\bar{\boldsymbol{P}}+n\boldsymbol{v}-\frac{1}{Le}\boldsymbol{\nabla} n\right)\cdot\boldsymbol{k}=0, \quad  \text{at} \quad x_3=1,
\end{align}
while for the rigid bottom surface
\begin{align}
\label{eqn:equation16}
    \boldsymbol{v}&=0, \quad \boldsymbol{v}\times\boldsymbol{k}=0, \quad  \mathcal{T}=1, \quad \nonumber\\
    &\left(\frac{1}{Le}n U_s\bar{\boldsymbol{P}}+n\boldsymbol{v}-\frac{1}{Le}\boldsymbol{\nabla} n\right)\cdot\boldsymbol{k}=0, \quad \text{at} \quad x_3=0.
\end{align}

Non-dimensional Radiative transfer equation (\ref{eqn:equation3}) becomes
\begin{equation}
    \frac{d I}{d r}+n \hbar  I(\boldsymbol{x},\boldsymbol{r})=0,
\end{equation}
where the optical depth of the suspension is $\hbar=\iota\bar{n}H$. 
The non-dimensional boundary condition for the top surface intensity becomes
\begin{equation}
    I(x_1,x_2,1,\alpha_1,\alpha_2)=I^0\delta(\boldsymbol{r}-\boldsymbol{r}^0) , \quad \pi/2\le\alpha_1\le\pi,
\end{equation}
while for the bottom surface intensity
\begin{equation}
    I(x_1,x_2,0,\alpha_1,\alpha_2)=0,\quad 0\le\alpha_1\le\pi/2.
\end{equation}

\section{Steady-state}
\label{sec3}
In the basic state, $\boldsymbol{v}=0$, $\mathcal{P}=\mathcal{P}_b$, $n=n_b(x_3)$, $I=I_b(x_3,\alpha_1)$ and $\mathcal{G}=\mathcal{G}_b(x_3)$ $\mathcal{T}=\mathcal{T}_b$. 

The radiative transfer equation at steady-state, after solving, yields
\begin{equation}
\label{eqn:equation18}
    \frac{\partial I_b}{\partial x_3}+\frac{\hbar n_b(x_3)}{\cos{\alpha_1}}I_b(x_3,\alpha_1)=0.
\end{equation}
The boundary condition for the radiative transfer equation is given by
\begin{equation}
\label{eqn:equation19}
    I_b(1,\alpha_1)=I^0\delta(\boldsymbol{r}-\boldsymbol{r}^0).
\end{equation}
The resolved form of the radiative transfer equation (\ref{eqn:equation18}) is

\begin{equation}
\label{eqn:equation20}
    I_b(x_3,\alpha_1)=C\exp{\left(\frac{-\hbar}{\cos{\alpha_1}}\int_1^{x_3} n_b(s)ds\right)},
\end{equation}
apply boundary condition (\ref{eqn:equation19}), 
\begin{equation}
\label{eqn:equation21}
    I_b(x_3,\alpha_1)=I^0\delta(\boldsymbol{r}-\boldsymbol{r}^0)\exp{\left(\frac{-\hbar}{\cos{\alpha_1}}\int_1^{x_3} n_b(s)ds\right)}.
\end{equation}
The expression for the total intensity $\mathcal{G}$ at the basic state is derived as

\begin{align}
\label{eqn:equation22}
    \mathcal{G}_b(x_3)&=\int_0^{4\pi} I_b(x_3,\alpha_1)d\omega =I^0\exp{\left(\hbar\int_1^{x_3} n_b(s)ds\right)}.
\end{align}

The basic state cell conservation equation transforms into
\begin{equation}
\label{eqn:equation24}
\frac{d n_b}{dx_3}-U_s T_b n_b=0,
\end{equation}
with 
\begin{equation}
\label{eqn:equation25}
    \int_0^1 n_b(x_3)dx_3=1.
\end{equation}

The expression for the basic state radiative heat flux becomes
\begin{align}
\label{eqn:equation26}
    \boldsymbol{q}_b(x_3)&=\int_0^{4\pi}I_b(x_3,\alpha_1)\boldsymbol{r}d\omega \nonumber \\
    &=-I^0\exp{\left(\hbar\int_1^{x_3} n_b(s)ds\right)\boldsymbol{k}} \nonumber\\
    &=|\boldsymbol{q}_b|(-\boldsymbol{k}).
\end{align}
Consequently, the mean swimming orientation $\bar{\boldsymbol{P}}_b$ at the basic state is calculated using

\begin{equation}
\label{eqn:equation27}
    \bar{\boldsymbol{P}}_b=-T(\mathcal{G}_b)\frac{\boldsymbol{q}_b}{|\boldsymbol{q}_b|}=T(\mathcal{G}_b)\boldsymbol{k}.
\end{equation}

Furthermore, the basic state temperature is
\begin{equation}
    \mathcal{T}_b=1-x_3.
\end{equation}

\section{Linear stability analysis}
\label{sec4}
To explore linear instability, an infinitesimal perturbation $\epsilon(0<\epsilon \ll 1)$ is applied in the basic state via

\begin{equation*}
    \boldsymbol{v}=\boldsymbol{0}+\epsilon \boldsymbol{v}^*(x_1,x_2,x_3,t)+O(\epsilon^2),\\
    \end{equation*}
    \begin{equation*}
    n=n_b(x_3)+\epsilon n^*(x_1,x_2,x_3,t)+O(\epsilon^2),\\
    \end{equation*}
    \begin{equation*}
    \mathcal{P}=\mathcal{P}_b+\epsilon \mathcal{P}^*+O(\epsilon^2),\\
    \end{equation*}
    \begin{equation*}
    \bar{\boldsymbol{P}}=\bar{\boldsymbol{P}}_b+\epsilon\bar{\boldsymbol{P}}^*+O(\epsilon^2),\\
    \end{equation*}
    \begin{equation*}
    \mathcal{G}=\mathcal{G}_b+\epsilon\mathcal{G^*}+O(\epsilon^2),\\
    \end{equation*}
    \begin{equation*}
    \mathcal{T}=\mathcal{T}_b+\epsilon T'+O(\epsilon^2),
    \end{equation*}

here perturbed fluid velocity is $\boldsymbol{v}^* = (u^* , v^* , w^*)$.
Thus, the linearized governing equations are written as
\begin{equation}
\label{eqn:equation28a}
    div(\boldsymbol{v}^*) =0,  
\end{equation}
\begin{equation}
\label{eqn:equation28b}
   \frac{1}{P_r}\frac{\partial \boldsymbol{v}^*}{\partial t}=\nabla^2\boldsymbol{v}^*-\boldsymbol{\nabla} \mathcal{P}^*-n^* R_a \boldsymbol{k}+R_T T' \boldsymbol{k},
\end{equation}
\begin{equation}
    \frac{\partial T'}{\partial t}-\boldsymbol{v}\cdot\boldsymbol{k}=\boldsymbol{\nabla}^2 T',
\end{equation}
\begin{equation}
\label{eqn:equation29}
    \frac{\partial n^*}{\partial t}=-\frac{dn_b}{dx_3}w^*+ \frac{1}{Le}\nabla^2 n^*-\frac{1}{Le}U_s\boldsymbol{\nabla}\cdot(n^*\bar{\boldsymbol{P}}_b+n_b\bar{\boldsymbol{P}}^*).
\end{equation}

 The total intensity can be written as
 \begin{align*}
     \mathcal{G}&=\mathcal{G}_b+\epsilon\mathcal{G^*}+O(\epsilon^2)\nonumber\\
     &=I^0\exp{\left(\hbar\int_1^{x_3}(n_b(s)+\epsilon n^*+O(\epsilon^2))ds\right)}.
 \end{align*}
On accumulating $O(\epsilon)$ terms, perturbed total intensity becomes
 \begin{equation*}
    \mathcal{G}^*=I^0\left(\hbar\int_1^{x_3} n^* ds\right)\exp{\left(\hbar\int_1^{x_3} n_b(s)ds\right)}.
\end{equation*}
The mean swimming orientation is given by
\begin{align}
\label{eqn:equation30}
    \bar{\boldsymbol{P}}&=\bar{\boldsymbol{P}}_b+\epsilon\bar{\boldsymbol{P}}^* +O(\epsilon^2) \nonumber \\
    &=T(\mathcal{G}_b+\epsilon \mathcal{G}^*+O(\epsilon^2))\boldsymbol{k}.
\end{align}
On accumulating $O(\epsilon)$ terms, perturbed mean swimming orientation becomes
\begin{equation}
\label{eqn:equation31}
    \bar{\boldsymbol{P}}^*=\mathcal{G}^*\frac{\partial T}{\partial \mathcal{G}_b}\boldsymbol{k}.
\end{equation}

Applying the curl operator twice to (\ref{eqn:equation28b}) and focusing on the vertical component yields 
\begin{align}
\label{eqn:equation34}
    \frac{1}{P_r}\frac{\partial}{\partial t}(\boldsymbol{\nabla}^2 w^*)&=\boldsymbol{\nabla}^4 w^*-R_a \boldsymbol{\nabla}^2 n^*+R_a\frac{\partial^2}{\partial x_3^2} n^*\nonumber\\
    &-R_T \boldsymbol{\nabla}^2 T'+R_T\frac{\partial^2}{\partial x_3^2} T'.
\end{align}

Also, taking the curl of equation (\ref{eqn:equation28b}), we get
\begin{equation}
\label{eqn:equation35}
    \frac{1}{P_r}\frac{\partial \zeta}{\partial t}-Ta^{1/2}\frac{\partial w^*}{\partial \bar{x}_3}=\boldsymbol{\nabla}^2 \zeta.
\end{equation}
Equation (\ref{eqn:equation35}) represents the vorticity equation and $\zeta$ is the vertical component of the vorticity.

In equation (\ref{eqn:equation29})
\begin{eqnarray}
\label{eqn:equation32}
    \boldsymbol{\nabla}\cdot(n^*\bar{\boldsymbol{P}}_b+n_b\bar{\boldsymbol{P}}^*)=2\hbar n_b \mathcal{G}_b\frac{d T(\mathcal{G}_b)}{d \mathcal{G}_b}+T(\mathcal{G}_b)\frac{\partial n^*}{\partial x_3} \nonumber \\
    +\hbar\frac{\partial}{\partial x_3}\left(n_b\mathcal{G}_b\frac{d T(\mathcal{G}_b)}{d\mathcal{G}_b}\right)\int_1^{x_3} n^* ds.
\end{eqnarray}
Therefore, equation (\ref{eqn:equation29}) re-write as
\begin{align}
\label{eqn:equation33}
    \frac{\partial n^*}{\partial t}-\frac{1}{Le}\boldsymbol{\nabla}^2 n^*-\frac{1}{Le}\hbar U_s\frac{\partial}{\partial x_3}\left(n_b\mathcal{G}_b\frac{d T(\mathcal{G}_b)}{d\mathcal{G}_b}\right)\int_{x_3}^1  n^* ds \nonumber \\
    +2\frac{1}{Le}\hbar n_b \mathcal{G}_b\frac{d T(\mathcal{G}_b)}{d \mathcal{G}_b}n^*U_s+\frac{1}{Le}T(\mathcal{G}_b)\frac{\partial n^*}{\partial x_3}U_s
    =-w^*\frac{d n_b}{d x_3}.
\end{align}

Perturbed boundary conditions for the stress-free top surface become
\begin{align*}
    w^*=0, \quad \frac{\partial^2 w^*}{\partial {x_3}^2}=0, \quad \frac{\partial \zeta}{\partial x_3}=0, \quad T'&=0,\quad U_s T_b n^*-\frac{\partial n^*}{\partial x_3}=0, \quad \text{at} \quad x_3=1,
\end{align*}
while for the rigid bottom surface
\begin{align*}
    w^*=0, \quad \frac{\partial w^*}{\partial x_3}=0,\quad \zeta=0, \quad T'=0, \quad \nonumber\\
    \hbar U_s n_b \mathcal{G}_b \frac{d T_b}{d \mathcal{G}_b} \int_{x_3}^1 n^* d \bar{x}_3-U_s T_b n^*+\frac{\partial n^*}{\partial x_3}=0 \quad 
    \text{at} \quad x_3=0.
\end{align*}

Normal modes are decomposed from linearized governing equations via 
\begin{equation*}
    w^*=\hat{w}(x_3)\exp{[\gamma t+i(a_1 x+a_2 y)]},
\end{equation*}
\begin{equation*}
   \zeta=Z(x_3)\exp{[\gamma t+i(a_1 x+a_2 y)]},
\end{equation*}
\begin{equation*}
   T'=\Theta(x_3)\exp{[\gamma t+i(a_1 x+a_2 y)]},
\end{equation*}
\begin{equation*}
    n^*=\hat{n}(x_3)\exp{[\gamma t+i(a_1 x+a_2 y)]}.
\end{equation*}

Here, $a_1$ and $a_2$ are wavenumbers in the $x_1$ and $x_2$ directions and the resultant $a=\sqrt{a_1^2+a_2^2}$ is a horizontal wavenumber.

Thus, the governing equations (\ref{eqn:equation28a})-(\ref{eqn:equation29}) in normal modes become 
\begin{eqnarray}
    \label{eqn:equation36}
    \frac{\gamma}{P_r}\left(\frac{d^2}{dx_3^2}-a^2\right) \hat{w}(x_3) -\left(\frac{d^2}{dx_3^2}-a^2\right)^2\hat{w}(x_3)
    = -Ta^{1/2}\frac{d Z(x_3)}{d x_3}+a^2R_a\hat{n}(x_3)-a^2R_T\Theta(x_3),
\end{eqnarray}
\begin{equation}
     \frac{\gamma}{P_r} Z(x_3)=Ta^{1/2}\frac{d \hat{w}(x_3)}{d x_3} + \left(\frac{d^2}{dz^2}-a^2\right) Z(x_3),
\end{equation}
\begin{equation}
    \label{eqn:equation37}
    \left(\gamma+a^2-\frac{d^2}{d x_3^2}\right)\Theta=\hat{w},
\end{equation}
\begin{eqnarray}
\label{eqn:equation38}
    &-\hbar U_s \frac{\partial}{\partial x_3}\left(n_b\mathcal{G}_b\frac{d T(\mathcal{G}_b)}{d\mathcal{G}_b}\right)\int_{x_3}^1  \hat{n}d \bar{x}_3+( \gamma Le+a^2) \hat{n}\nonumber \\
    &+2\hbar n_b \mathcal{G}_b\frac{d T_b}{d \mathcal{G}_b}U_s \hat{n}+U_s T_b \frac{d \hat{n}}{d x_3}-\frac{d^2 \hat{n}}{d x_3^2}
    =-Le \frac{d n_b}{d x_3}\hat{w}.
\end{eqnarray}

The associated boundary conditions in the normal modes for the stress-free top surface become
\begin{align*}
    \hat{w}(x_3)=0,\quad \frac{d^2\hat{w}(x_3)}{d x_3^2}=0,\quad \frac{d Z(x_3)}{d x_3}=0,\quad \Theta=0, \nonumber\\
    \quad U_s T_b \hat{n}-\frac{d \hat{n}}{d x_3}=0  \quad \text{at} \quad x_3=1,
\end{align*}
while for the rigid bottom surface
\begin{align*}
    \hat{w}(x_3)=0,\quad\frac{\hbar\hat{w}(x_3)}{d x_3}=0,\quad Z(x_3)=0,\quad \Theta=0,\quad \nonumber\\
    \hbar U_s n_b \mathcal{G}_b \frac{d T_b}{d \mathcal{G}_b} \int_{x_3}^1 \hat{n} d \bar{x}_3-U_s T_b \hat{n}+\frac{d \hat{n}}{d x_3}=0\quad 
    \text{at} \quad x_3=0.
\end{align*}

Define a new variable
\begin{equation}
\label{eqn:equation39}
    N(x_3)=\int_{x_3}^1 \hat{n} d \bar{x}_3,
\end{equation}
so that the system of equations becomes
\begin{align}
\label{eqn:equation40}
    \frac{d^4 \hat{w}}{d x_3}-\left(2 a^2+ \frac{\gamma}{P_r}\right)\frac{d^2 \hat{w}}{d x_3^2}+a^2\left(a^2+\frac{\gamma}{P_r}\right)\hat{w} \nonumber\\
    =Ta^{1/2}\frac{d Z}{d x_3}+a^2 R_a \frac{d N}{d x_3}+a^2 R_T \Theta,
\end{align}
\begin{equation}
    \frac{d^2 Z}{d x_3^2}-\left(a^2+\frac{\gamma}{P_r}\right)Z=-Ta^{1/2}\frac{d \hat{w}}{d x_3},
\end{equation}
\begin{equation}
\label{eqn:equation41}
    \left(\gamma+a^2-\frac{d^2}{d x_3^2}\right)\Theta=\hat{w},
\end{equation}
\begin{align}
\label{eqn:equation42}
    &\frac{d^3 N}{d x_3^3}-U_s T_b \frac{d^2 N}{d x_3^2}-\left(\gamma Le+a^2+2\hbar U_s n_b \mathcal{G}_b\frac{d T_b}{d \mathcal{G}_b}\right)\frac{d N}{d x_3} \nonumber \\
    &- \hbar U_s \frac{d}{d x_3}\left(n_b \mathcal{G}_b\frac{d T_b}{d \mathcal{G}_b}\right)N=-Le\frac{d n_b}{d x_3}\hat{w}.
\end{align}
Also, boundary conditions for the stress-free top surface become
\begin{align}
    \label{eqn:equation45}
    \hat{w}(x_3)=0,\quad \frac{d^2\hat{w}(x_3)}{d z^2}=0,\quad \frac{d Z(x_3)}{d x_3}=0 ,\quad \Theta=0, \quad \nonumber\\
    U_s T_b \frac{d N}{d x_3}-\frac{d^2 N}{d x_3^2}=0 \quad  \text{at} \quad x_3=1,
\end{align}
while for the rigid bottom surface
\begin{align}
    \label{eqn:equation43}
    \hat{w}(x_3)=0,\quad \frac{d\hat{w}(x_3)}{d x_3}=0,\quad Z(x_3)=0, \quad \Theta=0, \quad \nonumber\\
    \hbar U_s n_b \mathcal{G}_b \frac{d T_b}{d \mathcal{G}_b} N+U_s T_b \frac{d N}{d x_3}
    -\frac{d^2 N}{d x_3^2}=0\quad  \text{at} \quad x_3=0,
\end{align}

and 
\begin{equation}
\label{eqn:equation46}
    N(x_3)=0 \quad \text{at} \quad x_3=1.
\end{equation}

A set of governing equations and boundary conditions (\ref{eqn:equation40})-(\ref{eqn:equation46}) constitute a system of phototactic thermal-bioconvection equations. Without thermal convection in a non-rotating medium, the governing equations return to the form presented by \citet{ref9}. Also, without thermal convection in a rotating medium, the governing equations return to the form presented by \citet{kumar2023effect}.

\begin{table*}
\caption{Common suspension parameters for the phototactic microorganism \textit{Chlamydomonas}\cite{ref9,ref13,ref17,kumar2023effect}.}
\begin{tabular}{ p{8cm} p{5cm}}
\hline
\hline
Cell radius &$10^{-3}$cm\\
Ratio of cell density &$\Delta \varrho/\varrho=5\times 10^{-2}$ \\
Temperature difference& $\Delta T=1$K\\
Temperature of upper wall&$T_u^*=300$ K\\
Volumetric thermal expansion coefficient&$\beta=3.4\times 10^{-3}$K$^{-1}$\\
Thermal diffusivity&$\alpha_f=2\times 10^{-3}$cm$^2$/s\\
Cell diffusivity &$D=5\times10^{-5}-5\times10^{-4}$cm$^2$/s\\
Cell volume &$\vartheta=5\times10^{-10}$cm$^3$\\
Average cell swimming speed &$U_c =10^{-2}$cm/s\\
Average concentration &$\bar{n}=10^6$cm$^{-3}$\\
Prandtl number &$P_r=5$ \\
Kinematic viscosity &$\nu=10^{-2}$cm$^2$/s\\
Scaled average swimming speed &$U_s=20H$\\

\hline
\hline
\end{tabular}\\
\label{tab:table1}
\end{table*}

\section{Numerical results}
\label{sec5}
The set of ordinary differential equations (ODEs) represented by equations (\ref{eqn:equation40}) to (\ref{eqn:equation42}) constitutes an eleventh-order system that is accompanied by eleven boundary conditions (\ref{eqn:equation45}) to (\ref{eqn:equation46}). For dealing with coupled ODEs, a numerical method known as the shooting technique is used as part of the analysis approach. The MATLAB built-in bvp4c solver facilitates the computation \cite{shampine2003solving}. The investigation of the linear stability of the basic state will be the primary emphasis of this work. To do this, we will be illustrating neutral curves. These neutral curves are constituted by the points where the real component of the growth rate $Re(\gamma)$ equals zero. These zero points have relevance while evaluated in the context of stability analysis. The existence of an overstable or oscillatory solution is possible in cases if the imaginary component of the growth rate $Im(\gamma)$ along the neutral curve is not zero. On the other hand, according to the concept of exchange of stabilities, perturbations to the basic state are considered to be stationary if the value of $Im(\gamma)$ is equal to zero along this curve \cite{ref12}. In the time dependence which is expressed as $\exp(\gamma t) = \exp[(Re(\gamma)+i Im(\gamma))t]$, we include both the growth rate $Re(\gamma)$ as well as a hypothetical oscillation that is described by $Im(\gamma)/2\pi$. In situations with stable layering, the wave vectors have a negative value for the $Re(\gamma)$ parameter. However, as wave numbers venture just above the convective threshold, a narrow range comes into play where $Re(\gamma)$ becomes positive. The branch of the solution that exhibits the most significant amount of instability is the one in which the bioconvection Rayleigh number $R_a$ takes on its lowest possible value, which is represented by the symbol $R_a^c$. This critical solution, given the name $(a^c, R_a^c)$, is known for being the most unstable solution. To bring our model into line with the work that has been done in the past on phototactic bioconvection, we take into account phototactic microorganisms which are analogous to \textit{Chlamydomonas}. Consequently, we establish the required parameters for this study in reference to existing works like \citet{ref9,ref13,ref17, kumar2023effect} (refer to Table \ref{tab:table1}). The radiation parameters are the same as in Ref.\cite{ref14}. The optical depth can vary ranging from $0.25$ to $1$ for a suspension depth of $0.5$ cm, and $U_s = 10$ is the scaled swimming speed that corresponds to this range. Similarly, the value of $U_s$ for a suspension with a depth of $1.0$ cm is $20$. The Taylor numbers fall within the same range that is used in Ref. \cite{kumar2023effect}. To maintain compatibility with several other phototactic bioconvection models, $I^0 = 0.8$ will be kept constant throughout. It has been demonstrated that there is a connection between the critical intensity $\mathcal{G}_c$ and the parameter in such a way that $-1.1\le\chi\le 1.1$ results in $0.3\le\mathcal{G}_c\le0.8$ \cite{ref17,kumar2023effect}.

 \begin{table}
\caption{\label{tab:table2} Critical Rayleigh number and wavenumbers for fixed parameters $\hbar=0.5$, $U_s=10$, $Le=4$, $\mathcal{G}_c=0.68$, $\xi=0.19$, and $R_T=0$, $500$, $1100.5$.}
\begin{center}
\begin{tabular}{{p{1.95cm}p{1.95cm}p{1.95cm}p{1.95cm}p{1.95cm}p{1.95cm}p{1.95cm}}}
\hline\hline
&\multicolumn{2}{c}{$R_T=0$}&\multicolumn{2}{c}{$R_T=500$}&\multicolumn{2}{c}{$R_T=1100.5$}\\
\hline
$Ta$& $R_a^c$ & $a^c$& $R_a^c$ & $a^c$& $R_a^c$ & $a^c$\\
\hline
0 & 52.78 &1.6&31.74&2.3&0.01&2.7\\
100 & 62.9   & 2.1&39.18 &2.5&6.24 &2.8 \\
200 &71.14& 2.4 &45.88&2.7 & 12.09 &2.9\\
500   &90.97& 2.9&63.39 & 3.1&27.95& 3.2\\
1000 & 116.82& 3.4&87.56& 3.5& 50.79&3.6\\

\hline\hline
\end{tabular}
\end{center}
\end{table}

\begin{table}
\caption{\label{tab:table3} Critical Rayleigh number and wavenumbers for fixed parameters $\hbar=0.5$, $U_s=15$, $Le=4$, $\mathcal{G}_c=0.68$, $\xi=0.165$, and $Ta=0$, $800$.}
\begin{center}
\begin{tabular}{{p{1.9cm}p{1.9cm}p{1.9cm}p{1.9cm}p{1.9cm}p{1.9cm}p{1.9cm}p{1.9cm}}}
\hline\hline
\multicolumn{4}{c}{$Ta=0$}&\multicolumn{4}{c}{$Ta=800$}\\
\hline
$R_T$& $R_a^c$ & $a^c$& $Im(\gamma)$ & $R_T$&$R_a^c$& $a^c$ & $Im(\gamma)$\\
\hline
0 & 97.05 &3.3&0&0&133.54&4.1&0\\
100 & 89.21   & 3.3&0 &100&125.79 &4.1&0\\
200 &80.86& 3.4 &0&200& 177.61 &4.1&0\\
500   &52.41& 3.4&0 & 500&90.64& 4.2&0\\
800 & 17.71& 3.4&0& 1000& 34.41&4.3&0\\
931.2 & 0.09& 3.3&0& 1246& 0.07&4.3&0\\
\hline\hline
\end{tabular}
\end{center}
\end{table}

\begin{table}
\caption{\label{tab:table4}Critical Rayleigh number and wavenumbers for fixed parameters $\hbar=1.0$, $U_s=15$, $Le=8 $,$\mathcal{G}_c=0.51$, $\xi=-0.485$, and $Ta=0$ ,$400$, $2000$.}
\begin{center}
\begin{tabular}{p{1.2cm}p{1.2cm}p{1.2cm}p{1.2cm}p{1.2cm}p{1.2cm}p{1.2cm}p{1.2cm}p{1.2cm}p{1.2cm}p{1.2cm}p{1.2cm}}
\hline\hline
\multicolumn{4}{c}{$Ta=0$}&\multicolumn{4}{c}{$Ta=400$}&\multicolumn{4}{c}{$Ta=2000$}\\
\hline
$R_T$& $R_a^c$ & $a^c$& $Im(\gamma)$ & $R_T$&$R_a^c$& $a^c$ & $Im(\gamma)$& $R_T$&$R_a^c$& $a^c$ & $Im(\gamma)$\\
\hline
0 & 30.86 &2.0&11.54&0&55.23&2.6&14.04&0 & 97.72 &5.0&0\\
100 & 29.06   & 2.0 &11.15&100&53.38 &2.6&13.59&100 & 91.69  & 5.0 &0\\
200 &27.19& 2.0 &10.71&200& 51.54 &2.6&13.14&200 &85.61& 5.0 &0\\
500   &21.69& 2.0&9.38 & 500&41.61& 4.3&0&500   &67.1& 5.0&0\\
800 & 12.14& 3.8&0& 1000& 8.76&4.1&0&1000 & 35.25& 5.0&0\\
931.5 & 0.08& 3.3&0& 1115& 0.01&3.9&0&1528 & 0.08& 4.9&0\\
\hline\hline
\end{tabular}
\end{center}
\end{table}

\begin{figure*}
    \centering
    \includegraphics[width=16cm, height=12cm ]{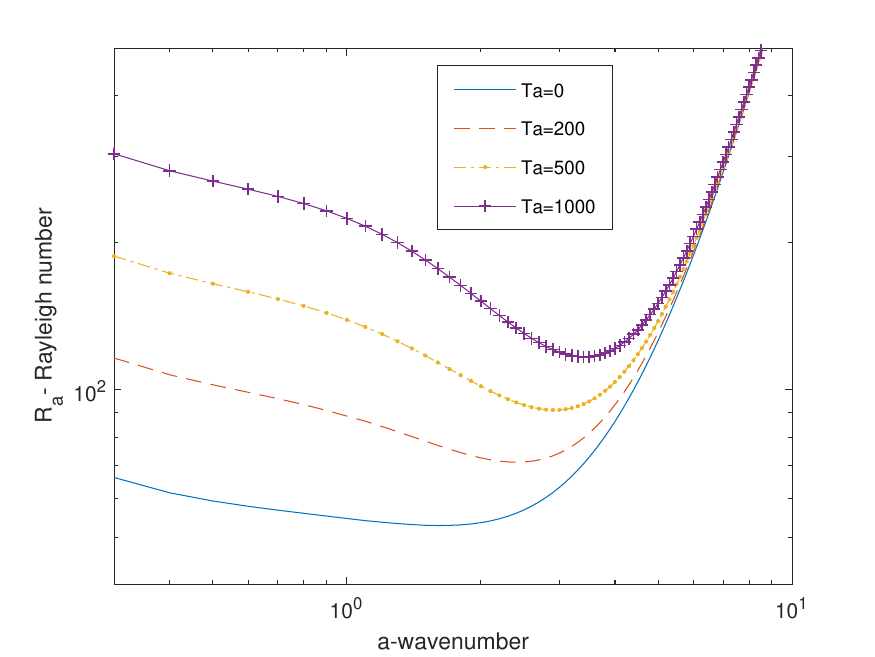}
    \caption{Neutral curves for fixed parameters $\hat{w}=10$, $d=0.5$, $\mathcal{G}_c=0.68$, $\xi=0.19$, $Le=4$, and $R_T=0$ as $Ta$ is varied.  }
   \label{fig:10v0.5k0.68Ic0Rt.pdf}
 \end{figure*}

As the Taylor number increases, neutral curves are described for some fixed parameters $\hat{w}=10$, $d=0.5$, $\mathcal{G}_c=0.68$, $\xi=0.19$, $Le=4$, and $R_T=0$ in Fig. \ref{fig:10v0.5k0.68Ic0Rt.pdf}. The critical wavenumber $a^c$ and Rayleigh number $R_a^c$ for $Ta=0$ are $1.6$ and $52.78$, respectively. The equation $\Lambda=2\pi/a$ allows us to calculate the wavelength $\Lambda$ of the initial disturbance. For $Ta=0$, the critical wavelength $\Lambda^c$ of a disturbance at the basic state is $3.92$, and the most unstable solution is found on a stationary branch. There are no oscillating branches shown as the Taylor number $Ta$ is increased; instead, the system solely displays stationary solutions. Rotation reduces vertical motion by restricting fluid movement to the horizontal plane, which in turn prevents bioconvection. As a result, rotation increases the stability of the system. This increased stability is reflected in the critical Rayleigh number $R_a^c$, which rises as the Taylor number $Ta$ increases. The critical Rayleigh number $R_a^c$, which grows as the Taylor number $Ta$ does, reflects this improved stability. The forced uniform rotation produces a Coriolis force and initiates the development of a vortex, which results in a dragging impact on the fluid. In practical terms, when the angular velocity increases or viscosity decreases, the fluid becomes more tightly connected to the vortex line, making it more challenging for movements perpendicular to the $\Omega$ to close the streamlines required for convection. This phenomenon explains how uniform rotation typically dampens instability \cite{chand1953}. 

As the thermal Rayleigh number $R_T$ is raised to $500$, the critical Rayleigh number $R_a^c$ decreases for all the Taylor numbers (see Table \ref{tab:table2}). In this case, the critical wavenumber $a^c$ and Rayleigh number $R_a^c$ for $Ta=0$ are $1.6$ and $52.78$, respectively. For $Ta=1000$, the critical Rayleigh number is $87.56$, and the critical wavenumber is $3.5$.

\begin{figure*}
    \centering
    \includegraphics[width=16cm, height=12cm ]{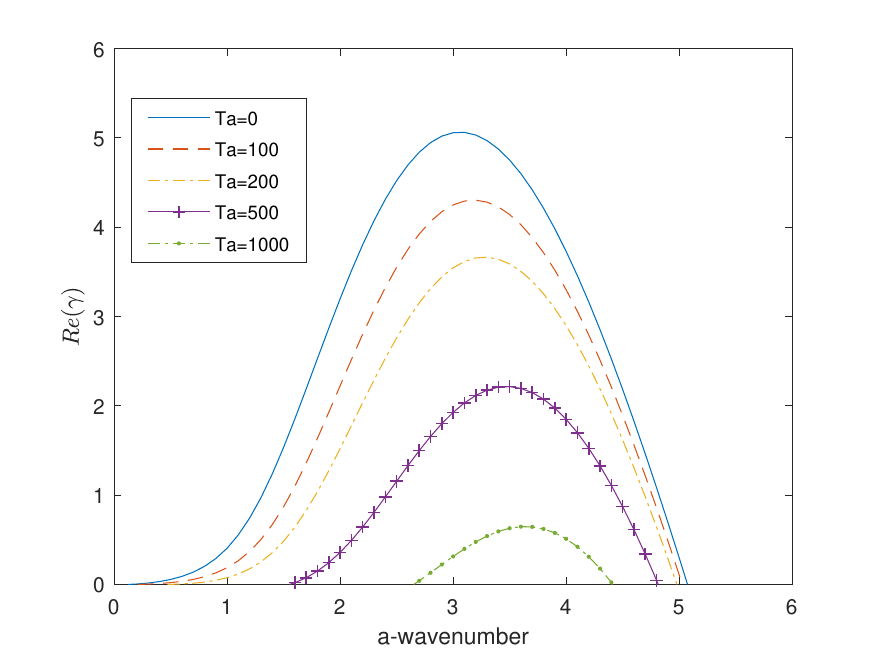}
    \caption{Growth rate for fixed parameters $\hat{w}=10$, $d=0.5$, $\mathcal{G}_c=0.68$, $\xi=0.19$, $Le=4$, $R_a=100$, and $R_T=500$ as $Ta$ is varied. }
   \label{fig:growthrate.pdf}
 \end{figure*}

One can indeed acquire estimations of the anticipated wavelengths of the primary perturbations by constructing growth rate curves for Rayleigh numbers exceeding the critical Rayleigh number, $R_a> R_a^c$. Figure \ref{fig:growthrate.pdf} depicts growth rate curves corresponding to different values of $Ta$ while other parameters $\hat{w}=10$, $d=0.5$, $\mathcal{G}_c=0.68$, $\xi=0.19$, $Le=4$, $R_a=100$, and $R_T=500$ are fixed. If $Ta=0$, the maximum growth rate $Re(\gamma)_{max}=5.06$ is observed at $a=3.0$, which corresponds to a pattern wavelength of $\Lambda=2.09$. As the value of $Ta$ increases, the wavelength of the pattern decreases even further. The maximum growth rate reaches $Re(\gamma)_{max}=0.65$ at $Ta = 1000$ and $a = 3.6$, which corresponds to the pattern's wavelength $\Lambda=1.74$. As the Taylor number further increases, the pattern wavelength approaches zero. This shows insights into how variations in the Taylor number influence pattern wavelengths within the system. 

\begin{figure*}
    \centering
    \includegraphics[width=16cm, height=12cm ]{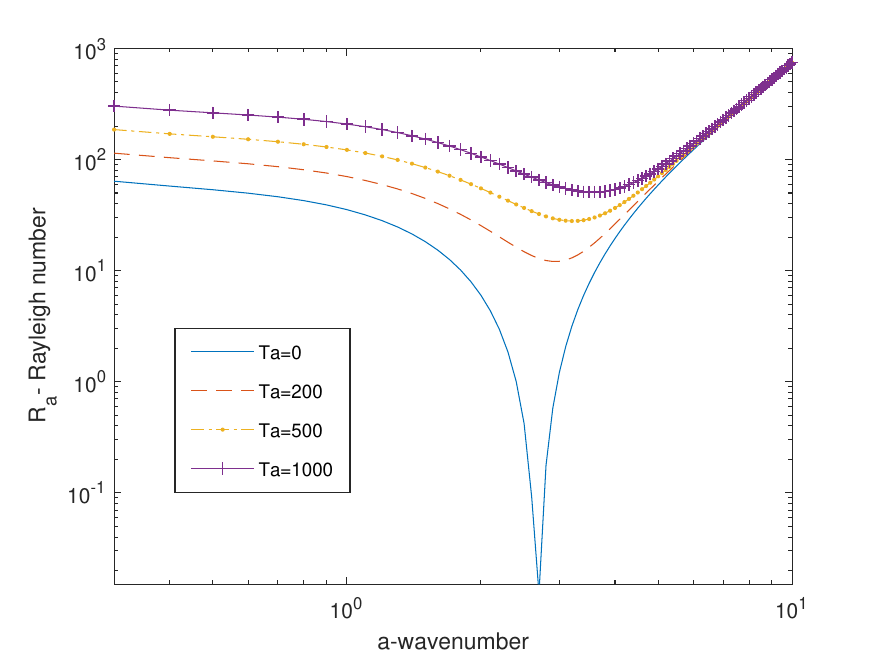}
    \caption{Neutral curves for fixed parameters $\hat{w}=10$, $d=0.5$, $\mathcal{G}_c=0.68$, $\xi=0.19$, $Le=4$, and $R_T=1100.5$ as $Ta$ is varied.  }
   \label{fig:10v0.5k0.68Ic1100.5Rt.pdf}
 \end{figure*}

 For $Ta=0$, the critical Rayleigh number approaches zero as $R_T$ approaches $1100.5$, and for other Taylor numbers, it decreases as well (see Fig. \ref{fig:10v0.5k0.68Ic1100.5Rt.pdf}). Therefore, if $Ta=0$, the suspension will become unstable on its own, and the bioconvection will be obscured by the traditional Rayleigh-B$\acute{e}$nared convection, even though certain bacteria may be able to tolerate the conditions.

\begin{figure*}
    \centering
    \includegraphics[width=16cm, height=12cm ]{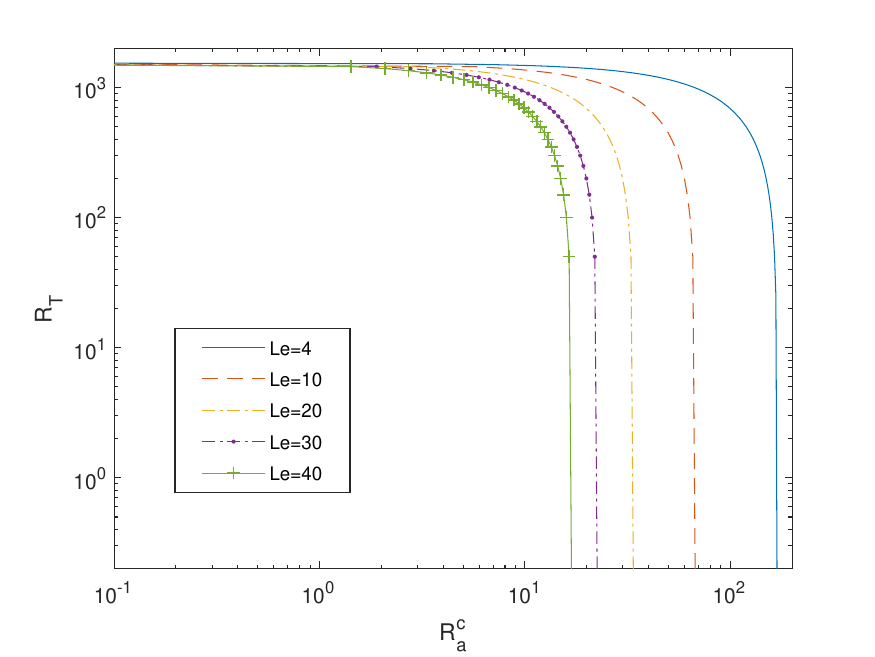}
    \caption{The relationship between $R_a^c$ and $R_T$ for $\hat{w}=10$, $d=1.0$, $\mathcal{G}_c=0.5$, $\xi=-0.5$ and $Ta=500$ as $Le$ is varied.  }
   \label{fig:10v1klewis.pdf}
 \end{figure*}

\begin{figure*}
    \centering
    \includegraphics[width=16cm, height=12cm ]{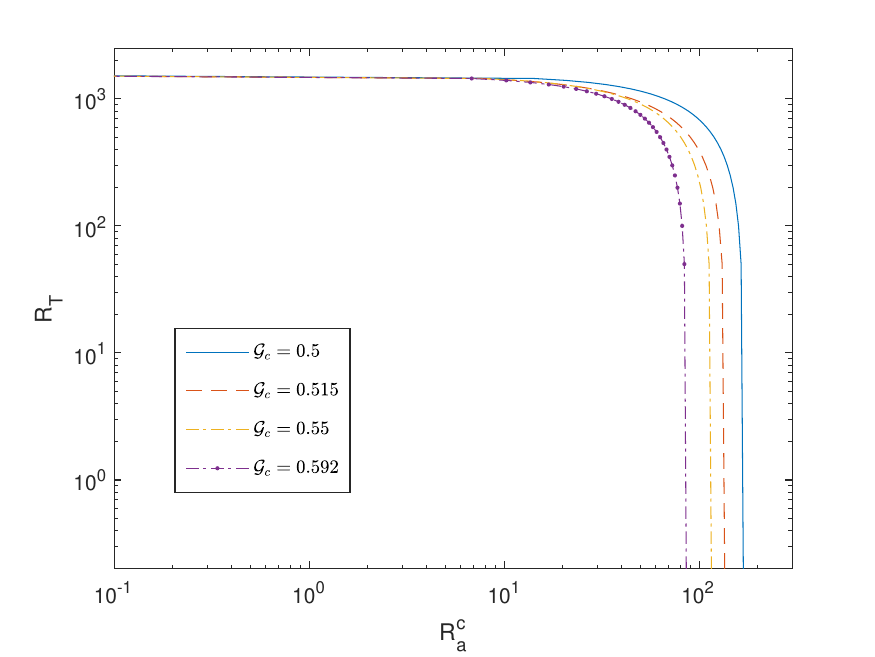}
    \caption{The relationship between $R_a^c$ and $R_T$ for $\hat{w}=10$, $d=1.0$, $Ta=500$, and $Le=4$ as $\mathcal{G}_c$ is varied.  }
   \label{fig:10v1kGc.pdf}
 \end{figure*}

\begin{figure*}
    \centering
    \includegraphics[width=16cm, height=12cm ]{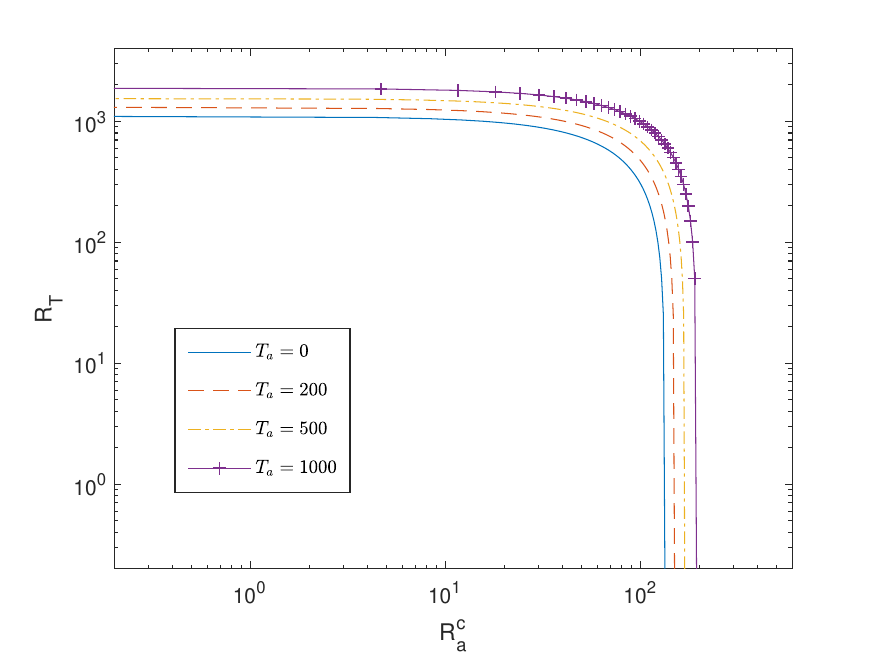}
    \caption{The relationship between $R_a^c$ and $R_T$ for $\hat{w}=10$, $d=1.0$, $\mathcal{G}_c=0.5$, $\xi=-0.5$ and $Le=4$ as $Ta$ is varied. }
   \label{fig:10v1keffRt.pdf}
 \end{figure*}

The correlation between $R_a^c$ and $R_T$ is explored in figures \ref{fig:10v1klewis.pdf}-\ref{fig:10v1keffRt.pdf} by manipulating $Ta$, $Le$, and $\mathcal{G}_c$, respectively. It has been observed that the $R_a^c$ decreases as the $R_T$ grows. The critical thermal Rayleigh number is $1547.1$, and it is the same for all different values of the Lewis number $Le$ under circumstances in which the other constant parameters are $\hat{w}=10$, $d=1.0$, $\mathcal{G}_c=0.5$, $\xi=-0.5$ and $Ta=500$ (see Fig. \ref{fig:10v1klewis.pdf}). From the data presented in Table (\ref{tab:table1}), it is evident that the diffusion coefficient $D$ ranges between $5\times 10 ^{-5} cm^2/s$ and $5\times 10 ^{-4} cm^2/s$, leading to a corresponding variation in the Lewis number $Le$ from $4$ to $40$. The graphical representation clearly indicates that an increase in the Lewis number $Le$ results in a simultaneous decrease in the critical Rayleigh number $R_a^c$. This observation implies that the thermal diffusion coefficient $D$ plays a stabilizing role within the system. The critical thermal Rayleigh number $R_T^c$ remains unaffected by changes in the Lewis number $Le$, representing its independence of this parameter. In contrast, the critical bioconvection Rayleigh number $R_a^c$ exhibits a notable dependence on the Lewis number $Le$. This characteristic can be directly linked to the Lewis number $Le$ concept, which is intimately associated with diffusivity. Interestingly, when $R_a=0$, the system's solution becomes decoupled from the diffusivity $D$, indicating that changes in diffusivity no longer exert a significant influence on the solution. Hence, it is reasonable to infer that a considerable portion of the bioconvection process is substantially impacted by the specific value of diffusivity. 

The critical thermal Rayleigh number is $1547.1$, and it is the same for all different values of the Lewis number $Le$ under circumstances in which the other constant parameters are $\hat{w}=10$, $d=1.0$, $Ta=500$, and $Le=4$ (see Fig. \ref{fig:10v1kGc.pdf}). The outcomes are similar to the preceding instance. The gravitational region below the sublayer grows with critical total intensity. This region promotes convection, resulting in the critical bioconvection Rayleigh number decreases with critical total intensity, rendering the system more unstable. Therefore, the critical thermal Rayleigh number $R_T^c$ is unaffected by the critical total intensity $\mathcal{G}_c$, however, the critical bioconvection Rayleigh number $R_a^c$ is significantly impacted by it.

The relationship between $R_a^c$ and $R_T$ for $\hat{w}=10$, $d=1.0$, $\mathcal{G}_c=0.5$, $\xi=-0.5$ and $Le=4$ as $Ta$ is varied. Increasing the Taylor number $Ta$ leads to an elevation of the critical bioconvection Rayleigh number $R_a^c$, which, in turn, enhances stability. As a result, we observe that the critical bioconvection Rayleigh number is significantly influenced by changes in the Taylor number. On the other hand, the critical thermal Rayleigh number increases with the Taylor number because higher Taylor numbers promote greater stability.

\begin{figure*}
    \centering
    \includegraphics[width=16cm, height=12cm ]{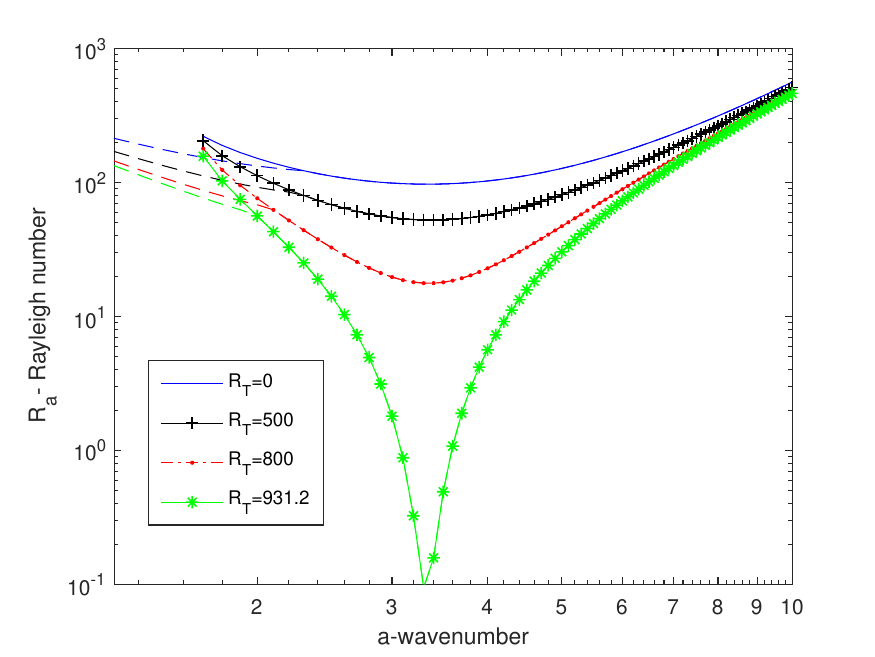}
    \caption{ Neutral curves for fixed parameters $\hat{w}=15$, $d=0.5$, $\mathcal{G}_c=0.68$, $\xi=0.165$, $Le=4$, and $Ta=0$ as $R_T$ is varied. }
   \label{fig:15v0.5k0Ta.pdf}
 \end{figure*}

\begin{figure*}
    \centering
    \includegraphics[width=16cm, height=12cm ]{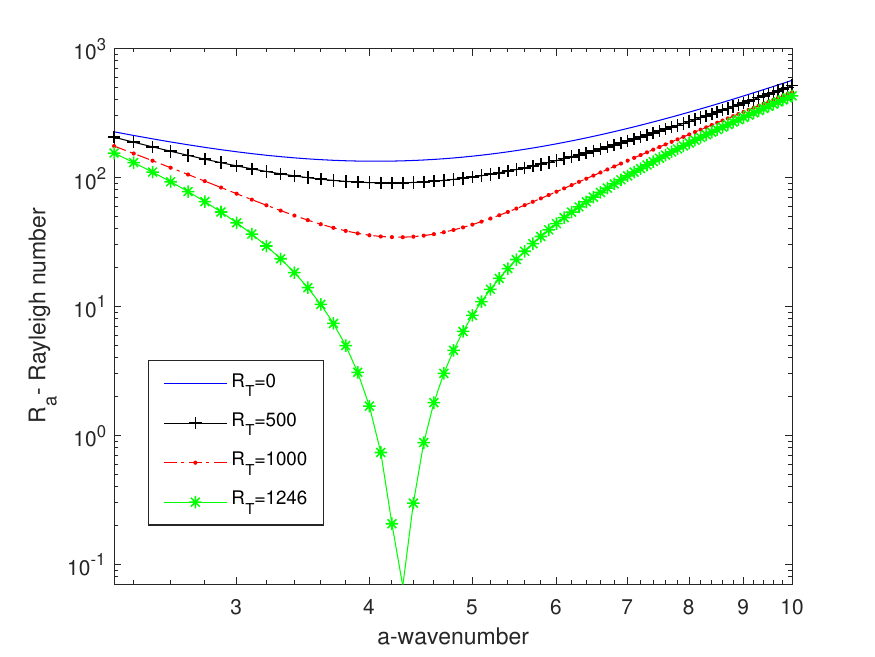}
    \caption{Neutral curves for fixed parameters $\hat{w}=15$, $d=0.5$, $\mathcal{G}_c=0.68$, $\xi=0.165$, $Le=4$, and $Ta=800$ as $R_T$ is varied. }
   \label{fig:15v0.5k800Ta.pdf}
 \end{figure*}

\begin{figure*}
    \centering
    \includegraphics[width=16cm, height=12cm ]{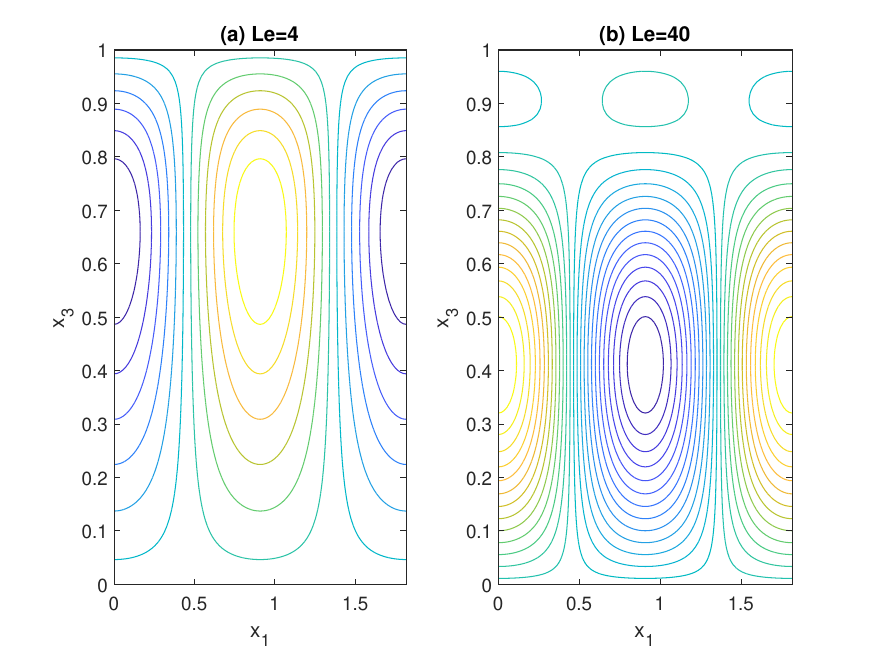}
    \caption{Flow patterns on the stationary branch for $\hat{w}=15$, $d=0.5$, $\mathcal{G}_c=0.68$, $\xi=0.165$, $R_T=0$, $Ta=800$, (a) $Le=4$, and (b) $Le=40$.  }
   \label{fig:phaseportrait.pdf}
 \end{figure*}

For the fixed parameters $\hat{w}=15$, $d=0.5$, $\mathcal{G}_c=0.68$, $\xi=0.165$, $Le=4$, and $Ta=0$, Fig. \ref{fig:15v0.5k0Ta.pdf} illustrates the linear stability curves for the variation of $R_T$. For $R_T=0$, an overstable branch deviates from the stationary branch at $R_a=116$, $a=2.4$, and the most unstable solution $(R_a^c, a^c)=(97.05,3.3)$ resides on the stationary branch (see Table \ref{tab:table3}). With an increase in $R_T$, the critical bioconvection Rayleigh number decreases. The impact of $R_T$ is most prominently observed in the vicinity of the most unstable solution. In this scenario, as $R_T$ approaches $931.2$, the critical Rayleigh number approaches zero. In this scenario, an overstable branch persists, and the most unstable solution continues to exist on the stationary branch. As the Taylor number $Ta$ increases to $800$, we find that the overstable branch vanishes, and only stationary curves are observed (see Fig. \ref{fig:15v0.5k800Ta.pdf}). The perturbed temperature flow pattern is depicted in the Fig. \ref{fig:phaseportrait.pdf} for fixed parameters: $\hat{w}=15$, $d=0.5$, $\mathcal{G}_c=0.68$, $\xi=0.165$, $R_T=0$, $Ta=800$, (a) $Le=4$, and (b) $Le=40$. Mode $n$ is a solution in which $n$ convection cells are stacked vertically one over the other. In the case when $Le=4$, the bioconvection solution consists of a single convection cell throughout the suspension. As a result, the bioconvective solution exhibits characteristics of the mode $1$ type at $Le=4$. With an increase in the Lewis number $Le$, the initially single bioconvection cell, extending through the entire depth of the layer, undergoes a transformation. It becomes accompanied by a secondary, smaller convection cell that originates at the upper portion of the layer. This secondary cell then expands in height as $Le$ continues to increase. Consequently, in Fig. \ref{fig:phaseportrait.pdf}(b), which corresponds to a Lewis number of $40$, the bioconvection solution is depicted with two convection cells.

For the fixed parameters $\hat{w}=15$, $d=1.0$, $\mathcal{G}_c=0.51$, $\xi=-0.485$, $Le=8$, and $Ta=0$, Fig. \ref{fig:15v0.5k0Ta.pdf} illustrates the linear stability curves for the variation of $R_T$. At $Ta=0$, an oscillatory/overstable branch separates from the stationary branch at a wavenumber of $a=3.9$. In this scenario, we observe that the most unstable solution resides on the oscillatory branch. The critical Rayleigh number is $30.86$, corresponding to the critical wavenumber of $a^c=2$. When $R_T$ is increased to $800$, the oscillatory branch remains present, although the most unstable solution shifts to the stable branch. Therefore, as the Taylor number increases, a transition can be identified from the most unstable mode in the oscillatory state to the stationary state. As $R_T$ is increased to $931.5$, the critical Rayleigh number approaches zero, causing the system itself to become unstable. 

\begin{figure}
    \centering
    \includegraphics[width=16cm, height=20cm ]{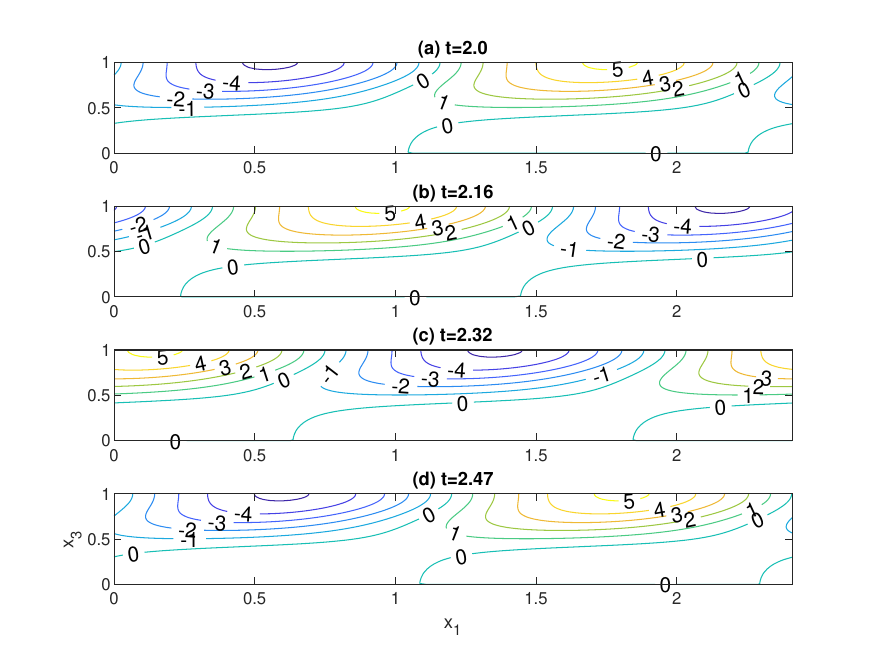}
    \caption{Flow pattern produced by the perturbed vorticity $\zeta$	 throughout one cycle of oscillation for  $\hat{w}=15$, $d=1.0$, $\mathcal{G}_c=0.51$, $\xi=-0.485$, $Le=8$, $R_T=200$, $R_a^c=51.54$, $a_c=2.6$, $Im(\gamma)=13.14$, (a) $t=2.0$, (b) $t=2.16$, (c) $t=2.32$, and (d) $t=2.47$.}
   \label{fig:periodicTEMP.pdf}
 \end{figure}

By increasing $Ta$ to $400$, the most stable solution remains on the overstable branch up to $R_T=200$. In this scenario, overstability occurs at $a_c=2.6$ and $R_a^c=51.54$ (see Table \ref{tab:table4}). A pair of complex conjugate eigenvalues, $\gamma=0\pm 13.14$, corresponds to $R_T=200$. This change is known as a Hopf bifurcation. The bioconvective flow patterns are mirror reflections of each other, corresponding to the complex conjugate pair of eigenvalues. One oscillation cycle takes $2\pi/{Im(\gamma)}=0.47$ units of time to complete. The bioconvective fluid movements become nonlinear in a much shorter time frame than the initially projected duration in an unstable state. Consequently, it is possible to observe the convection cells and flow patterns that occur throughout a single cycle of oscillation by using the perturbed eigenmode $\zeta$. Figures \ref{fig:periodicTEMP.pdf} illustrate the flow patterns generated by the perturbed vorticity $\zeta$ during one oscillation cycle. The fixed parameters include $\hat{w}=15$, $d=1.0$, $\mathcal{G}_c=0.51$, $\xi=-0.485$, $Le=8$, $R_T=200$, $R_a^c=51.54$, $a_c=2.6$, $Im(\gamma)=13.14$, (a) $t=2.0$, (b) $t=2.16$, (c) $t=2.32$, and (d) $t=2.47$. The wavelength of the most unstable solution is $\Lambda^c=3.14$. It shows the solution shift to the left, indicating the solution for the traveling wave. Increasing $Ta$ further to $2000$ results in the critical Rayleigh number for $R_T=0$ increasing to $97.72$. An oscillatory branch separates from the stationary branch at a wavenumber of $a=3.7$. However, the most unstable solution shifts to the stationary branch. When $R_T$ is increased to $500$, the oscillatory branch disappears, leaving only the stationary branch. As $R_T$ is further increased, only the stationary branch is observed, and at $R_T=1500$, the critical Rayleigh number approaches $0$.

\section{Conclusion}
\label{sec6}

This study is focused on the linear stability analysis of a suspension that contains phototactic microorganisms and is heated from below in a rotating medium. Our analysis revealed that the interplay of parameters such as the Taylor number $Ta$, thermal Rayleigh number $R_T$, Lewis number $Le$, and critical total intensity $\mathcal{G}_c$ profoundly impacts the stability of the system. Our results highlighted how changes in these parameters influenced critical Rayleigh numbers and wavelengths, ultimately determining whether the system exhibited oscillatory or stationary behavior. Increasing the critical total intensity $\mathcal{G}_c$ has a destabilizing effect on the system. This effect suggests that stronger light intensity can lead to more pronounced bioconvection patterns. It is observed that as the Taylor number increases due to the Coriolis force dampening vertical motion and promoting stationary solutions, the system tends towards a more stable state. Higher thermal Rayleigh numbers tend to decrease the critical bioconvection Rayleigh number, rendering the system more prone to instability. Furthermore, the significance of the Lewis number was highlighted, as variations in the thermal diffusion coefficient were found to strongly influence the critical Rayleigh number. This emphasizes the critical role of diffusivity in the bioconvection process. It also explored the fascinating phenomenon of transitions between stationary and oscillatory states with the variation of the Taylor number and thermal Rayleigh number. These transitions, characterized by bifurcations and oscillatory behavior, provided deeper insights into the complex nature of the bioconvection patterns. The region beneath the sublayer, which is gravitationally unstable, fosters convection, while the region above it inhibits it. Furthermore, phototaxis can either facilitate or impede the convection taking place in the suspension. Consequently, the existence of oscillatory solutions is attributed to the interplay between stabilizing and destabilizing processes \cite{ref10}.

\section{Acknowledgements}
The University Grants Commission financially supports this study, Grants number 191620003662, New Delhi (India).

\section{Data availability}
The data that support the findings of this study are in the article. 
\section{Declaration of Interests}
The authors state that they have no conflicting interests.

\nocite{*}
\section{Reference}
\bibliography{aipsamp}

\providecommand{\noopsort}[1]{}\providecommand{\singleletter}[1]{#1}%
\begin{thebibliography}{40}%
\makeatletter
\providecommand \@ifxundefined [1]{%
 \@ifx{#1\undefined}
}%
\providecommand \@ifnum [1]{%
 \ifnum #1\expandafter \@firstoftwo
 \else \expandafter \@secondoftwo
 \fi
}%
\providecommand \@ifx [1]{%
 \ifx #1\expandafter \@firstoftwo
 \else \expandafter \@secondoftwo
 \fi
}%
\providecommand \natexlab [1]{#1}%
\providecommand \enquote  [1]{``#1''}%
\providecommand \bibnamefont  [1]{#1}%
\providecommand \bibfnamefont [1]{#1}%
\providecommand \citenamefont [1]{#1}%
\providecommand \href@noop [0]{\@secondoftwo}%
\providecommand \href [0]{\begingroup \@sanitize@url \@href}%
\providecommand \@href[1]{\@@startlink{#1}\@@href}%
\providecommand \@@href[1]{\endgroup#1\@@endlink}%
\providecommand \@sanitize@url [0]{\catcode `\\12\catcode `\$12\catcode `\&12\catcode `\#12\catcode `\^12\catcode `\_12\catcode `\%12\relax}%
\providecommand \@@startlink[1]{}%
\providecommand \@@endlink[0]{}%
\providecommand \url  [0]{\begingroup\@sanitize@url \@url }%
\providecommand \@url [1]{\endgroup\@href {#1}{\urlprefix }}%
\providecommand \urlprefix  [0]{URL }%
\providecommand \Eprint [0]{\href }%
\providecommand \doibase [0]{http://dx.doi.org/}%
\providecommand \selectlanguage [0]{\@gobble}%
\providecommand \bibinfo  [0]{\@secondoftwo}%
\providecommand \bibfield  [0]{\@secondoftwo}%
\providecommand \translation [1]{[#1]}%
\providecommand \BibitemOpen [0]{}%
\providecommand \bibitemStop [0]{}%
\providecommand \bibitemNoStop [0]{.\EOS\space}%
\providecommand \EOS [0]{\spacefactor3000\relax}%
\providecommand \BibitemShut  [1]{\csname bibitem#1\endcsname}%
\let\auto@bib@innerbib\@empty
\bibitem [{\citenamefont {Pedley}\ and\ \citenamefont {Kessler}(1992)}]{ref1}%
  \BibitemOpen
  \bibfield  {author} {\bibinfo {author} {\bibfnamefont {T.~J.}\ \bibnamefont {Pedley}}\ and\ \bibinfo {author} {\bibfnamefont {J.~O.}\ \bibnamefont {Kessler}},\ }\bibfield  {title} {\enquote {\bibinfo {title} {Hydrodynamic phenomena in suspensions of swimming micro-organisms},}\ }\href@noop {} {\bibfield  {journal} {\bibinfo  {journal} {Annu. Rev. Fluid Mech.}\ }\textbf {\bibinfo {volume} {24}},\ \bibinfo {pages} {313--358} (\bibinfo {year} {1992})}\BibitemShut {NoStop}%
\bibitem [{\citenamefont {Wager}(1911)}]{ref2}%
  \BibitemOpen
  \bibfield  {author} {\bibinfo {author} {\bibfnamefont {H.~W.~T.}\ \bibnamefont {Wager}},\ }\bibfield  {title} {\enquote {\bibinfo {title} {On the effect of gravity upon the movements and aggregation of euglena viridis, ehrb., and other micro-organisms},}\ }\href@noop {} {\bibfield  {journal} {\bibinfo  {journal} {Phil. Trans. R. Soc. Lond. B}\ }\textbf {\bibinfo {volume} {201}},\ \bibinfo {pages} {333--390} (\bibinfo {year} {1911})}\BibitemShut {NoStop}%
\bibitem [{\citenamefont {Platt}(1961)}]{ref3}%
  \BibitemOpen
  \bibfield  {author} {\bibinfo {author} {\bibfnamefont {J.~R.}\ \bibnamefont {Platt}},\ }\bibfield  {title} {\enquote {\bibinfo {title} {‘‘bioconvection patterns’’ in cultures of free-swimming organisms},}\ }\href@noop {} {\bibfield  {journal} {\bibinfo  {journal} {Science}\ }\textbf {\bibinfo {volume} {133}},\ \bibinfo {pages} {1766--1767} (\bibinfo {year} {1961})}\BibitemShut {NoStop}%
\bibitem [{\citenamefont {Brinkmann}(1968)}]{ref6}%
  \BibitemOpen
  \bibfield  {author} {\bibinfo {author} {\bibfnamefont {K.}~\bibnamefont {Brinkmann}},\ }\bibfield  {title} {\enquote {\bibinfo {title} {An phasengrenzen induzierte ein und zweidimensionale kristallmuster in kulturen von euglena gracilis},}\ }\href@noop {} {\bibfield  {journal} {\bibinfo  {journal} {Z. Pflanzen Physiol.}\ }\textbf {\bibinfo {volume} {59}},\ \bibinfo {pages} {364--376} (\bibinfo {year} {1968})}\BibitemShut {NoStop}%
\bibitem [{\citenamefont {Kessler}(985b)}]{ref5}%
  \BibitemOpen
  \bibfield  {author} {\bibinfo {author} {\bibfnamefont {J.~O.}\ \bibnamefont {Kessler}},\ }\bibfield  {title} {\enquote {\bibinfo {title} {Co-operative and concentrative phenomena of swimming microorganisms},}\ }\href@noop {} {\bibfield  {journal} {\bibinfo  {journal} {Contemp. Phys.}\ }\textbf {\bibinfo {volume} {26}},\ \bibinfo {pages} {147--166} (\bibinfo {year} {1985b})}\BibitemShut {NoStop}%
\bibitem [{\citenamefont {Nultsch}\ and\ \citenamefont {Hoff}(1993)}]{ref4}%
  \BibitemOpen
  \bibfield  {author} {\bibinfo {author} {\bibfnamefont {W.}~\bibnamefont {Nultsch}}\ and\ \bibinfo {author} {\bibfnamefont {E.}~\bibnamefont {Hoff}},\ }\bibfield  {title} {\enquote {\bibinfo {title} {Investigations on pattern formatin in euglenae},}\ }\href@noop {} {\bibfield  {journal} {\bibinfo  {journal} {Arch. Protistenk}\ }\textbf {\bibinfo {volume} {115}},\ \bibinfo {pages} {336--352} (\bibinfo {year} {1993})}\BibitemShut {NoStop}%
\bibitem [{\citenamefont {Williams}\ and\ \citenamefont {Bees}(2011)}]{ref7}%
  \BibitemOpen
  \bibfield  {author} {\bibinfo {author} {\bibfnamefont {C.~R.}\ \bibnamefont {Williams}}\ and\ \bibinfo {author} {\bibfnamefont {M.~A.}\ \bibnamefont {Bees}},\ }\bibfield  {title} {\enquote {\bibinfo {title} {A tale of three taxes: Photo-gyro-gravitactic bioconvection},}\ }\href@noop {} {\bibfield  {journal} {\bibinfo  {journal} {J. Exp. Biol.}\ }\textbf {\bibinfo {volume} {214}},\ \bibinfo {pages} {2398--2408} (\bibinfo {year} {2011})}\BibitemShut {NoStop}%
\bibitem [{\citenamefont {Häder}(1987)}]{ref8}%
  \BibitemOpen
  \bibfield  {author} {\bibinfo {author} {\bibfnamefont {D.~P.}\ \bibnamefont {Häder}},\ }\bibfield  {title} {\enquote {\bibinfo {title} {Polarotaxis, gravitaxis and vertical phototaxis in the green flagellate, euglena gracilis},}\ }\href@noop {} {\bibfield  {journal} {\bibinfo  {journal} {Arch. Microbiol.}\ }\textbf {\bibinfo {volume} {147}},\ \bibinfo {pages} {179--183} (\bibinfo {year} {1987})}\BibitemShut {NoStop}%
\bibitem [{\citenamefont {Straughan}(1993)}]{ref10}%
  \BibitemOpen
  \bibfield  {author} {\bibinfo {author} {\bibfnamefont {B.}~\bibnamefont {Straughan}},\ }\enquote {\bibinfo {title} {\textit{Mathematical aspects of penetrative convection}},}\ \ (\bibinfo  {publisher} {Longman Scientific},\ \bibinfo {address} {New York},\ \bibinfo {year} {1993})\BibitemShut {NoStop}%
\bibitem [{\citenamefont {Greenspan}(1968)}]{ref11}%
  \BibitemOpen
  \bibfield  {author} {\bibinfo {author} {\bibfnamefont {H.~P.}\ \bibnamefont {Greenspan}},\ }\enquote {\bibinfo {title} {\textit{The theory of rotating fluids}},}\ \ (\bibinfo  {publisher} {Cambridge University Press},\ \bibinfo {address} {London},\ \bibinfo {year} {1968})\BibitemShut {NoStop}%
\bibitem [{\citenamefont {Chandrasekhar}(1961)}]{ref12}%
  \BibitemOpen
  \bibfield  {author} {\bibinfo {author} {\bibfnamefont {S.}~\bibnamefont {Chandrasekhar}},\ }\enquote {\bibinfo {title} {\textit{Hydrodynamic and hydromagnetic stability}},}\ \ (\bibinfo  {publisher} {Oxford University Press},\ \bibinfo {year} {1961})\BibitemShut {NoStop}%
\bibitem [{\citenamefont {Sekar}\ and\ \citenamefont {Vaidyanathan}(1993)}]{ref12-1}%
  \BibitemOpen
  \bibfield  {author} {\bibinfo {author} {\bibfnamefont {R.}~\bibnamefont {Sekar}}\ and\ \bibinfo {author} {\bibfnamefont {G.}~\bibnamefont {Vaidyanathan}},\ }\bibfield  {title} {\enquote {\bibinfo {title} {Convective instability of a magnetized ferrofluid in a rotating porous medium},}\ }\href@noop {} {\bibfield  {journal} {\bibinfo  {journal} {International journal of engineering science}\ }\textbf {\bibinfo {volume} {31}},\ \bibinfo {pages} {1139--1150} (\bibinfo {year} {1993})}\BibitemShut {NoStop}%
\bibitem [{\citenamefont {Venkatasubramanian}\ and\ \citenamefont {Kaloni}(1994)}]{ref12-2}%
  \BibitemOpen
  \bibfield  {author} {\bibinfo {author} {\bibfnamefont {S.}~\bibnamefont {Venkatasubramanian}}\ and\ \bibinfo {author} {\bibfnamefont {P.}~\bibnamefont {Kaloni}},\ }\bibfield  {title} {\enquote {\bibinfo {title} {Effects of rotation on the thermoconvective instability of a horizontal layer of ferrofluids},}\ }\href@noop {} {\bibfield  {journal} {\bibinfo  {journal} {International journal of engineering science}\ }\textbf {\bibinfo {volume} {32}},\ \bibinfo {pages} {237--256} (\bibinfo {year} {1994})}\BibitemShut {NoStop}%
\bibitem [{\citenamefont {Auernhammer}\ and\ \citenamefont {Brand}(2000)}]{ref12-3}%
  \BibitemOpen
  \bibfield  {author} {\bibinfo {author} {\bibfnamefont {G.}~\bibnamefont {Auernhammer}}\ and\ \bibinfo {author} {\bibfnamefont {H.}~\bibnamefont {Brand}},\ }\bibfield  {title} {\enquote {\bibinfo {title} {Thermal convection in a rotating layer of a magnetic fluid},}\ }\href@noop {} {\bibfield  {journal} {\bibinfo  {journal} {The European Physical Journal B-Condensed Matter and Complex Systems}\ }\textbf {\bibinfo {volume} {16}},\ \bibinfo {pages} {157--168} (\bibinfo {year} {2000})}\BibitemShut {NoStop}%
\bibitem [{\citenamefont {Ruo}, \citenamefont {Chang},\ and\ \citenamefont {Chen}(2010)}]{ref12-4}%
  \BibitemOpen
  \bibfield  {author} {\bibinfo {author} {\bibfnamefont {A.-C.}\ \bibnamefont {Ruo}}, \bibinfo {author} {\bibfnamefont {M.-H.}\ \bibnamefont {Chang}}, \ and\ \bibinfo {author} {\bibfnamefont {F.}~\bibnamefont {Chen}},\ }\bibfield  {title} {\enquote {\bibinfo {title} {Effect of rotation on the electrohydrodynamic instability of a fluid layer with an electrical conductivity gradient},}\ }\href@noop {} {\bibfield  {journal} {\bibinfo  {journal} {Physics of Fluids}\ }\textbf {\bibinfo {volume} {22}},\ \bibinfo {pages} {024102} (\bibinfo {year} {2010})}\BibitemShut {NoStop}%
\bibitem [{\citenamefont {Yadav}, \citenamefont {Agrawal},\ and\ \citenamefont {Bhargava}(2011)}]{ref12-5}%
  \BibitemOpen
  \bibfield  {author} {\bibinfo {author} {\bibfnamefont {D.}~\bibnamefont {Yadav}}, \bibinfo {author} {\bibfnamefont {G.}~\bibnamefont {Agrawal}}, \ and\ \bibinfo {author} {\bibfnamefont {R.}~\bibnamefont {Bhargava}},\ }\bibfield  {title} {\enquote {\bibinfo {title} {Thermal instability of rotating nanofluid layer},}\ }\href@noop {} {\bibfield  {journal} {\bibinfo  {journal} {International Journal of Engineering Science}\ }\textbf {\bibinfo {volume} {49}},\ \bibinfo {pages} {1171--1184} (\bibinfo {year} {2011})}\BibitemShut {NoStop}%
\bibitem [{\citenamefont {Mahajan}\ and\ \citenamefont {Arora}(2013)}]{ref12-6}%
  \BibitemOpen
  \bibfield  {author} {\bibinfo {author} {\bibfnamefont {A.}~\bibnamefont {Mahajan}}\ and\ \bibinfo {author} {\bibfnamefont {M.}~\bibnamefont {Arora}},\ }\bibfield  {title} {\enquote {\bibinfo {title} {Convection in rotating magnetic nanofluids},}\ }\href@noop {} {\bibfield  {journal} {\bibinfo  {journal} {Applied Mathematics and Computation}\ }\textbf {\bibinfo {volume} {219}},\ \bibinfo {pages} {6284--6296} (\bibinfo {year} {2013})}\BibitemShut {NoStop}%
\bibitem [{\citenamefont {Arain}\ \emph {et~al.}(2021)\citenamefont {Arain}, \citenamefont {Bhatti}, \citenamefont {Zeeshan},\ and\ \citenamefont {Alzahrani}}]{ref12-7}%
  \BibitemOpen
  \bibfield  {author} {\bibinfo {author} {\bibfnamefont {M.~B.}\ \bibnamefont {Arain}}, \bibinfo {author} {\bibfnamefont {M.~M.}\ \bibnamefont {Bhatti}}, \bibinfo {author} {\bibfnamefont {A.}~\bibnamefont {Zeeshan}}, \ and\ \bibinfo {author} {\bibfnamefont {F.~S.}\ \bibnamefont {Alzahrani}},\ }\bibfield  {title} {\enquote {\bibinfo {title} {Bioconvection reiner-rivlin nanofluid flow between rotating circular plates with induced magnetic effects, activation energy and squeezing phenomena},}\ }\href@noop {} {\bibfield  {journal} {\bibinfo  {journal} {Mathematics}\ }\textbf {\bibinfo {volume} {9}},\ \bibinfo {pages} {2139} (\bibinfo {year} {2021})}\BibitemShut {NoStop}%
\bibitem [{\citenamefont {Waqas}\ \emph {et~al.}(2021)\citenamefont {Waqas}, \citenamefont {Naseem}, \citenamefont {Muhammad},\ and\ \citenamefont {Farooq}}]{waqas}%
  \BibitemOpen
  \bibfield  {author} {\bibinfo {author} {\bibfnamefont {H.}~\bibnamefont {Waqas}}, \bibinfo {author} {\bibfnamefont {R.}~\bibnamefont {Naseem}}, \bibinfo {author} {\bibfnamefont {T.}~\bibnamefont {Muhammad}}, \ and\ \bibinfo {author} {\bibfnamefont {U.}~\bibnamefont {Farooq}},\ }\bibfield  {title} {\enquote {\bibinfo {title} {Bioconvection flow of casson nanofluid by rotating disk with motile microorganisms},}\ }\href@noop {} {\bibfield  {journal} {\bibinfo  {journal} {Journal of Materials Research and Technology}\ }\textbf {\bibinfo {volume} {13}},\ \bibinfo {pages} {2392--2407} (\bibinfo {year} {2021})}\BibitemShut {NoStop}%
\bibitem [{\citenamefont {Kuznetsov}(2005)}]{kuznetsov2005onset}%
  \BibitemOpen
  \bibfield  {author} {\bibinfo {author} {\bibfnamefont {A.}~\bibnamefont {Kuznetsov}},\ }\bibfield  {title} {\enquote {\bibinfo {title} {The onset of bioconvection in a suspension of gyrotactic microorganisms in a fluid layer of finite depth heated from below},}\ }\href@noop {} {\bibfield  {journal} {\bibinfo  {journal} {International Communications in Heat and Mass Transfer}\ }\textbf {\bibinfo {volume} {32}},\ \bibinfo {pages} {574--582} (\bibinfo {year} {2005})}\BibitemShut {NoStop}%
\bibitem [{\citenamefont {Kuznetsov}(2011)}]{kuznetsov2011non}%
  \BibitemOpen
  \bibfield  {author} {\bibinfo {author} {\bibfnamefont {A.}~\bibnamefont {Kuznetsov}},\ }\bibfield  {title} {\enquote {\bibinfo {title} {Non-oscillatory and oscillatory nanofluid bio-thermal convection in a horizontal layer of finite depth},}\ }\href@noop {} {\bibfield  {journal} {\bibinfo  {journal} {European Journal of Mechanics-B/Fluids}\ }\textbf {\bibinfo {volume} {30}},\ \bibinfo {pages} {156--165} (\bibinfo {year} {2011})}\BibitemShut {NoStop}%
\bibitem [{\citenamefont {Zhao}, \citenamefont {Xiao},\ and\ \citenamefont {Wang}(2018)}]{zhao2018linear}%
  \BibitemOpen
  \bibfield  {author} {\bibinfo {author} {\bibfnamefont {M.}~\bibnamefont {Zhao}}, \bibinfo {author} {\bibfnamefont {Y.}~\bibnamefont {Xiao}}, \ and\ \bibinfo {author} {\bibfnamefont {S.}~\bibnamefont {Wang}},\ }\bibfield  {title} {\enquote {\bibinfo {title} {Linear stability of thermal-bioconvection in a suspension of gyrotactic micro-organisms},}\ }\href@noop {} {\bibfield  {journal} {\bibinfo  {journal} {International Journal of Heat and Mass Transfer}\ }\textbf {\bibinfo {volume} {126}},\ \bibinfo {pages} {95--102} (\bibinfo {year} {2018})}\BibitemShut {NoStop}%
\bibitem [{\citenamefont {Zhao}\ \emph {et~al.}(2019)\citenamefont {Zhao}, \citenamefont {Wang}, \citenamefont {Wang},\ and\ \citenamefont {Mahabaleshwar}}]{zhao2019darcy}%
  \BibitemOpen
  \bibfield  {author} {\bibinfo {author} {\bibfnamefont {M.}~\bibnamefont {Zhao}}, \bibinfo {author} {\bibfnamefont {S.}~\bibnamefont {Wang}}, \bibinfo {author} {\bibfnamefont {H.}~\bibnamefont {Wang}}, \ and\ \bibinfo {author} {\bibfnamefont {U.}~\bibnamefont {Mahabaleshwar}},\ }\bibfield  {title} {\enquote {\bibinfo {title} {Darcy--brinkman bio-thermal convection in a suspension of gyrotactic microorganisms in a porous medium},}\ }\href@noop {} {\bibfield  {journal} {\bibinfo  {journal} {Neural Computing and Applications}\ }\textbf {\bibinfo {volume} {31}},\ \bibinfo {pages} {1061--1067} (\bibinfo {year} {2019})}\BibitemShut {NoStop}%
\bibitem [{\citenamefont {Balla}\ \emph {et~al.}(2020)\citenamefont {Balla}, \citenamefont {Ramesh}, \citenamefont {Kishan}, \citenamefont {Rashad},\ and\ \citenamefont {Abdelrahman}}]{balla2020bioconvection}%
  \BibitemOpen
  \bibfield  {author} {\bibinfo {author} {\bibfnamefont {C.~S.}\ \bibnamefont {Balla}}, \bibinfo {author} {\bibfnamefont {A.}~\bibnamefont {Ramesh}}, \bibinfo {author} {\bibfnamefont {N.}~\bibnamefont {Kishan}}, \bibinfo {author} {\bibfnamefont {A.}~\bibnamefont {Rashad}}, \ and\ \bibinfo {author} {\bibfnamefont {Z.}~\bibnamefont {Abdelrahman}},\ }\bibfield  {title} {\enquote {\bibinfo {title} {Bioconvection in oxytactic microorganism-saturated porous square enclosure with thermal radiation impact},}\ }\href@noop {} {\bibfield  {journal} {\bibinfo  {journal} {Journal of Thermal Analysis and Calorimetry}\ }\textbf {\bibinfo {volume} {140}},\ \bibinfo {pages} {2387--2395} (\bibinfo {year} {2020})}\BibitemShut {NoStop}%
\bibitem [{\citenamefont {Hussain}, \citenamefont {Raizah},\ and\ \citenamefont {Aly}(2022)}]{hussain2022thermal}%
  \BibitemOpen
  \bibfield  {author} {\bibinfo {author} {\bibfnamefont {S.}~\bibnamefont {Hussain}}, \bibinfo {author} {\bibfnamefont {Z.}~\bibnamefont {Raizah}}, \ and\ \bibinfo {author} {\bibfnamefont {A.~M.}\ \bibnamefont {Aly}},\ }\bibfield  {title} {\enquote {\bibinfo {title} {Thermal radiation impact on bioconvection flow of nano-enhanced phase change materials and oxytactic microorganisms inside a vertical wavy porous cavity},}\ }\href@noop {} {\bibfield  {journal} {\bibinfo  {journal} {International Communications in Heat and Mass Transfer}\ }\textbf {\bibinfo {volume} {139}},\ \bibinfo {pages} {106454} (\bibinfo {year} {2022})}\BibitemShut {NoStop}%
\bibitem [{\citenamefont {Vincent}\ and\ \citenamefont {Hill}(1996)}]{ref9}%
  \BibitemOpen
  \bibfield  {author} {\bibinfo {author} {\bibfnamefont {R.~V.}\ \bibnamefont {Vincent}}\ and\ \bibinfo {author} {\bibfnamefont {N.~A.}\ \bibnamefont {Hill}},\ }\bibfield  {title} {\enquote {\bibinfo {title} {Bioconvection in a suspension of phototactic algae},}\ }\href@noop {} {\bibfield  {journal} {\bibinfo  {journal} {J. Fluid Mech.}\ }\textbf {\bibinfo {volume} {327}},\ \bibinfo {pages} {343--371} (\bibinfo {year} {1996})}\BibitemShut {NoStop}%
\bibitem [{\citenamefont {Ghorai}\ and\ \citenamefont {Hill}(2005)}]{ref13}%
  \BibitemOpen
  \bibfield  {author} {\bibinfo {author} {\bibfnamefont {S.}~\bibnamefont {Ghorai}}\ and\ \bibinfo {author} {\bibfnamefont {N.~A.}\ \bibnamefont {Hill}},\ }\bibfield  {title} {\enquote {\bibinfo {title} {Penetrative phototactic bioconvection},}\ }\href@noop {} {\bibfield  {journal} {\bibinfo  {journal} {Phys. Fluids}\ }\textbf {\bibinfo {volume} {17}},\ \bibinfo {pages} {074101} (\bibinfo {year} {2005})}\BibitemShut {NoStop}%
\bibitem [{\citenamefont {Ghorai}, \citenamefont {Panda},\ and\ \citenamefont {Hill}(2010)}]{ref14}%
  \BibitemOpen
  \bibfield  {author} {\bibinfo {author} {\bibfnamefont {S.}~\bibnamefont {Ghorai}}, \bibinfo {author} {\bibfnamefont {M.~K.}\ \bibnamefont {Panda}}, \ and\ \bibinfo {author} {\bibfnamefont {N.~A.}\ \bibnamefont {Hill}},\ }\bibfield  {title} {\enquote {\bibinfo {title} {Bioconvection in a suspension of isotropically scattering phototactic algae},}\ }\href@noop {} {\bibfield  {journal} {\bibinfo  {journal} {Phys. Fluids}\ }\textbf {\bibinfo {volume} {22}},\ \bibinfo {pages} {071901} (\bibinfo {year} {2010})}\BibitemShut {NoStop}%
\bibitem [{\citenamefont {Kumar}(2023{\natexlab{a}})}]{kumar2023}%
  \BibitemOpen
  \bibfield  {author} {\bibinfo {author} {\bibfnamefont {S.}~\bibnamefont {Kumar}},\ }\bibfield  {title} {\enquote {\bibinfo {title} {Isotropic scattering with a rigid upper surface at the onset of phototactic bioconvection},}\ }\href@noop {} {\bibfield  {journal} {\bibinfo  {journal} {Physics of Fluids}\ }\textbf {\bibinfo {volume} {35}},\ \bibinfo {pages} {024106} (\bibinfo {year} {2023}{\natexlab{a}})}\BibitemShut {NoStop}%
\bibitem [{\citenamefont {Ghorai}\ and\ \citenamefont {Panda}(2013)}]{ref15}%
  \BibitemOpen
  \bibfield  {author} {\bibinfo {author} {\bibfnamefont {S.}~\bibnamefont {Ghorai}}\ and\ \bibinfo {author} {\bibfnamefont {M.~K.}\ \bibnamefont {Panda}},\ }\bibfield  {title} {\enquote {\bibinfo {title} {Bioconvection in an anisotropic scattering suspension of phototactic algae},}\ }\href@noop {} {\bibfield  {journal} {\bibinfo  {journal} {Eur. J. Mech.-B/Fluids}\ }\textbf {\bibinfo {volume} {41}},\ \bibinfo {pages} {81--93} (\bibinfo {year} {2013})}\BibitemShut {NoStop}%
\bibitem [{\citenamefont {Panda}\ \emph {et~al.}(2016)\citenamefont {Panda}, \citenamefont {Singh}, \citenamefont {Mishra},\ and\ \citenamefont {Mohanty}}]{ref16}%
  \BibitemOpen
  \bibfield  {author} {\bibinfo {author} {\bibfnamefont {M.~K.}\ \bibnamefont {Panda}}, \bibinfo {author} {\bibfnamefont {R.}~\bibnamefont {Singh}}, \bibinfo {author} {\bibfnamefont {A.~C.}\ \bibnamefont {Mishra}}, \ and\ \bibinfo {author} {\bibfnamefont {S.~K.}\ \bibnamefont {Mohanty}},\ }\bibfield  {title} {\enquote {\bibinfo {title} {Effects of both diffuse and collimated incident radiation on phototactic bioconvection},}\ }\href@noop {} {\bibfield  {journal} {\bibinfo  {journal} {Phys. Fluids}\ }\textbf {\bibinfo {volume} {28}},\ \bibinfo {pages} {124104} (\bibinfo {year} {2016})}\BibitemShut {NoStop}%
\bibitem [{\citenamefont {Panda}(2020)}]{panda2020effects}%
  \BibitemOpen
  \bibfield  {author} {\bibinfo {author} {\bibfnamefont {M.}~\bibnamefont {Panda}},\ }\bibfield  {title} {\enquote {\bibinfo {title} {Effects of anisotropic scattering on the onset of phototactic bioconvection with diffuse and collimated irradiation},}\ }\href@noop {} {\bibfield  {journal} {\bibinfo  {journal} {Physics of Fluids}\ }\textbf {\bibinfo {volume} {32}} (\bibinfo {year} {2020})}\BibitemShut {NoStop}%
\bibitem [{\citenamefont {Panda}, \citenamefont {Sharma},\ and\ \citenamefont {Kumar}(2022)}]{ref17}%
  \BibitemOpen
  \bibfield  {author} {\bibinfo {author} {\bibfnamefont {M.~K.}\ \bibnamefont {Panda}}, \bibinfo {author} {\bibfnamefont {P.}~\bibnamefont {Sharma}}, \ and\ \bibinfo {author} {\bibfnamefont {S.}~\bibnamefont {Kumar}},\ }\bibfield  {title} {\enquote {\bibinfo {title} {Effect of oblique irradiation on the onset of phototactic bioconvection},}\ }\href@noop {} {\bibfield  {journal} {\bibinfo  {journal} {Phys. Fluids}\ }\textbf {\bibinfo {volume} {34}},\ \bibinfo {pages} {024108} (\bibinfo {year} {2022})}\BibitemShut {NoStop}%
\bibitem [{\citenamefont {Kumar}(2022)}]{ref18}%
  \BibitemOpen
  \bibfield  {author} {\bibinfo {author} {\bibfnamefont {S.}~\bibnamefont {Kumar}},\ }\bibfield  {title} {\enquote {\bibinfo {title} {Phototactic isotropic scattering bioconvection with oblique irradiation},}\ }\href@noop {} {\bibfield  {journal} {\bibinfo  {journal} {Phys. Fluids}\ }\textbf {\bibinfo {volume} {34}},\ \bibinfo {pages} {114125} (\bibinfo {year} {2022})}\BibitemShut {NoStop}%
\bibitem [{\citenamefont {Panda}\ and\ \citenamefont {Rajput}(2023)}]{panda2023phototactic}%
  \BibitemOpen
  \bibfield  {author} {\bibinfo {author} {\bibfnamefont {M.}~\bibnamefont {Panda}}\ and\ \bibinfo {author} {\bibfnamefont {S.}~\bibnamefont {Rajput}},\ }\bibfield  {title} {\enquote {\bibinfo {title} {Phototactic bioconvection with the combined effect of diffuse and oblique collimated flux on an algal suspension},}\ }\href@noop {} {\bibfield  {journal} {\bibinfo  {journal} {Physics of Fluids}\ }\textbf {\bibinfo {volume} {35}} (\bibinfo {year} {2023})}\BibitemShut {NoStop}%
\bibitem [{\citenamefont {Kumar}(2023{\natexlab{b}})}]{kumar2023effect}%
  \BibitemOpen
  \bibfield  {author} {\bibinfo {author} {\bibfnamefont {S.}~\bibnamefont {Kumar}},\ }\bibfield  {title} {\enquote {\bibinfo {title} {Effect of rotation on the suspension of phototactic bioconvection},}\ }\href@noop {} {\bibfield  {journal} {\bibinfo  {journal} {Physics of Fluids}\ }\textbf {\bibinfo {volume} {35}} (\bibinfo {year} {2023}{\natexlab{b}})}\BibitemShut {NoStop}%
\bibitem [{\citenamefont {Modest}(2003)}]{ref-modest}%
  \BibitemOpen
  \bibfield  {author} {\bibinfo {author} {\bibfnamefont {M.~F.}\ \bibnamefont {Modest}},\ }\enquote {\bibinfo {title} {\textit{Radiative Heat Transfer}},}\ \ (\bibinfo  {publisher} {Academic Press},\ \bibinfo {address} {New York},\ \bibinfo {year} {2003})\ \bibinfo {edition} {2nd}\ ed.\BibitemShut {Stop}%
\bibitem [{\citenamefont {Chandrasekhar}(1960)}]{ref-chand}%
  \BibitemOpen
  \bibfield  {author} {\bibinfo {author} {\bibfnamefont {S.}~\bibnamefont {Chandrasekhar}},\ }\enquote {\bibinfo {title} {\textit{Radiative Transfer}},}\ \ (\bibinfo  {publisher} {Dover},\ \bibinfo {address} {New York},\ \bibinfo {year} {1960})\BibitemShut {NoStop}%
\bibitem [{\citenamefont {Shampine}, \citenamefont {Gladwell},\ and\ \citenamefont {Thompson}(2003)}]{shampine2003solving}%
  \BibitemOpen
  \bibfield  {author} {\bibinfo {author} {\bibfnamefont {L.~F.}\ \bibnamefont {Shampine}}, \bibinfo {author} {\bibfnamefont {I.}~\bibnamefont {Gladwell}}, \ and\ \bibinfo {author} {\bibfnamefont {S.}~\bibnamefont {Thompson}},\ }\href@noop {} {\emph {\bibinfo {title} {Solving ODEs with matlab}}}\ (\bibinfo  {publisher} {Cambridge university press},\ \bibinfo {year} {2003})\BibitemShut {NoStop}%
\bibitem [{\citenamefont {Chandrasekhar}(1953)}]{chand1953}%
  \BibitemOpen
  \bibfield  {author} {\bibinfo {author} {\bibfnamefont {S.}~\bibnamefont {Chandrasekhar}},\ }\bibfield  {title} {\enquote {\bibinfo {title} {The instability of a layer of fluid heated below and subject to coriolis forces},}\ }\href@noop {} {\bibfield  {journal} {\bibinfo  {journal} {Proceedings of the Royal Society of London. Series A. Mathematical and Physical Sciences}\ }\textbf {\bibinfo {volume} {217}},\ \bibinfo {pages} {306--327} (\bibinfo {year} {1953})}\BibitemShut {NoStop}%
\end{thebibliography}%


\begin{figure*}
    \centering
    \includegraphics[width=16cm, height=12cm ]{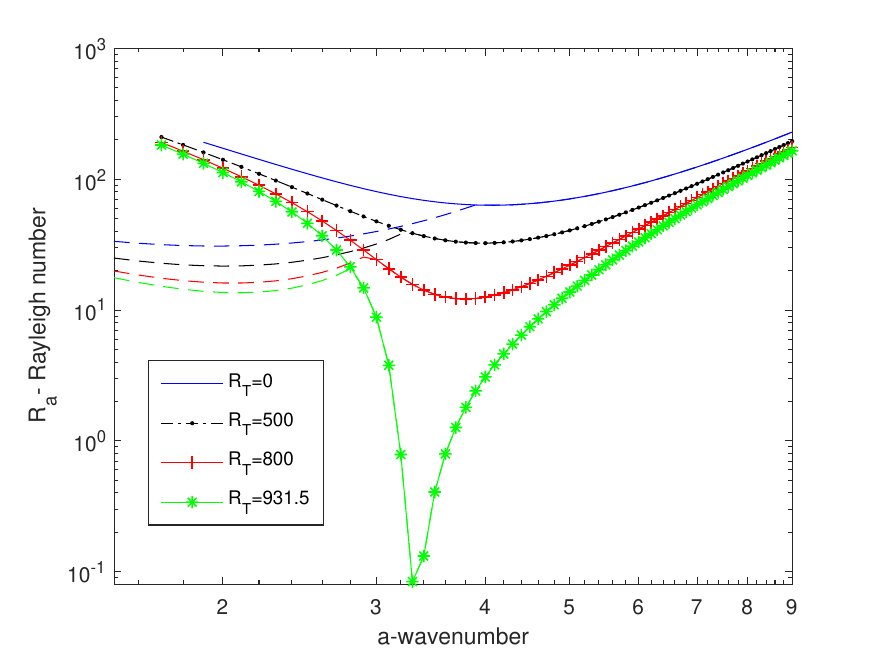}
    \caption{ Neutral curves for fixed parameters $\hat{w}=15$, $d=1.0$, $\mathcal{G}_c=0.51$, $\xi=-0.485$, $Le=8$, and $Ta=0$ as $R_T$ is varied. }
   \label{fig:15v1k0Ta.pdf}
 \end{figure*}
\begin{figure}
    \centering
    \includegraphics[width=16cm, height=12cm ]{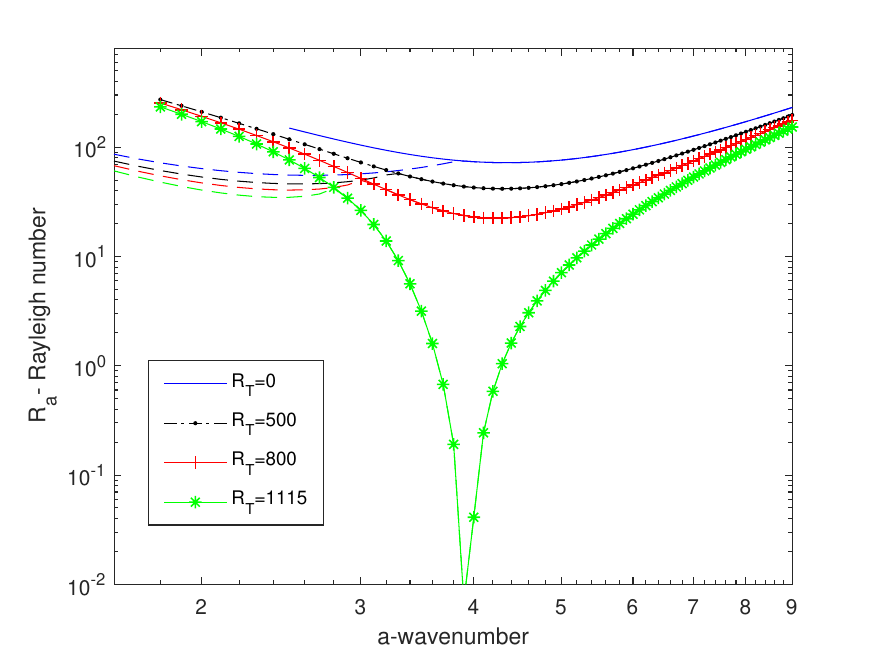}
    \caption{Neutral curves for fixed parameters $\hat{w}=15$, $d=1.0$, $\mathcal{G}_c=0.51$, $\xi=-0.485$, $Le=8$, and $Ta=400$ as $R_T$ is varied. }
   \label{fig:15v1k400Ta.pdf}
 \end{figure}

\begin{figure}
    \centering
    \includegraphics[width=16cm, height=11cm ]{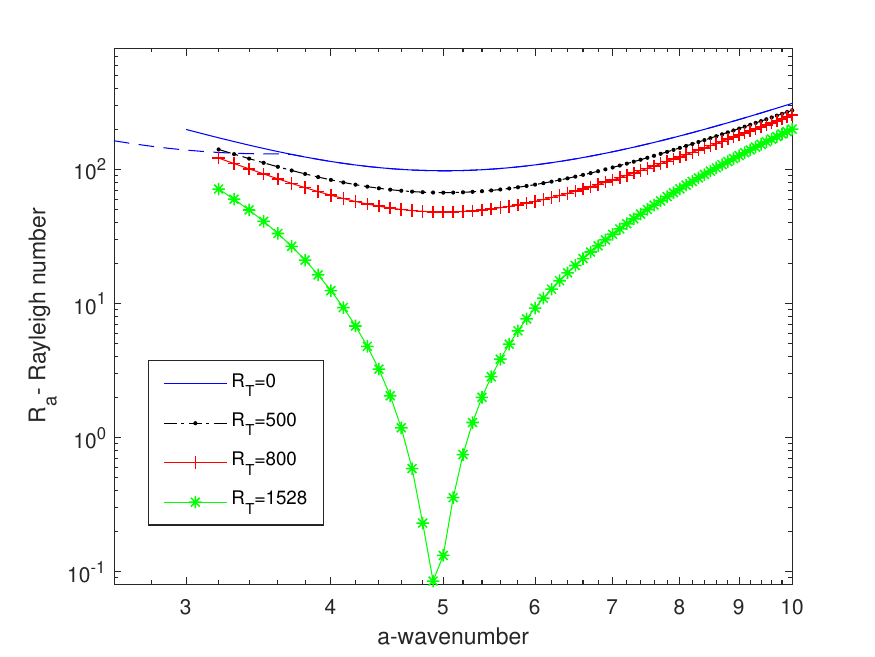}
    \caption{Neutral curves for fixed parameters $\hat{w}=15$, $d=1.0$, $\mathcal{G}_c=0.51$, $\xi=-0.485$, $Le=8$, and $Ta=2000$ as $R_T$ is varied. }
   \label{fig:15v1k2000Ta.pdf}
 \end{figure}

\end{document}